\documentclass[sigconf]{acmart}

\AtBeginDocument{%
  \providecommand\BibTeX{{%
    \normalfont B\kern-0.5em{\scshape i\kern-0.25em b}\kern-0.8em\TeX}}}

\copyrightyear{2025}
\acmYear{2025}
\setcopyright{rightsretained}
\acmConference[CHI '25]{CHI Conference on Human Factors in Computing
Systems}{April 26-May 1, 2025}{Yokohama, Japan}
\acmBooktitle{CHI Conference on Human Factors in Computing Systems (CHI
'25), April 26-May 1, 2025, Yokohama,
Japan}\acmDOI{10.1145/3706598.3713348}
\acmISBN{979-8-4007-1394-1/25/04}

\usepackage{algorithm}
\usepackage{mathtools}
\usepackage{algpseudocode}
\usepackage{graphicx}
\usepackage{color,soul}
\usepackage{comment}
\usepackage{soul}

\begin{document}

\title[CARING-AI]{CARING-AI: Towards Authoring Context-aware Augmented Reality INstruction through Generative Artificial Intelligence}
\author{Jingyu Shi}
\authornote{Three authors contributed equally to this research.}
\email{shi537@purdue.edu}
\orcid{0000-0001-5159-2235}
\affiliation{%
  \institution{Purdue University}
  \streetaddress{585 Purdue Mall}
  \city{West Lafayette}
  \state{Indiana}
  \country{USA}
  \postcode{47907}
}

\author{Rahul Jain}
\authornotemark[1]
\email{jain348@purdue.edu}
\orcid{0009-0001-3723-5482}
\affiliation{%
  \institution{Purdue University}
  \streetaddress{585 Purdue Mall}
  \city{West Lafayette}
  \state{Indiana}
  \country{USA}
  \postcode{47907}
}

\author{Seungguen Chi}
\authornotemark[1]
\email{chi65@purdue.edu}
\orcid{0000-0001-6965-6938}
\affiliation{%
  \institution{Purdue University}
  \streetaddress{585 Purdue Mall}
  \city{West Lafayette}
  \state{Indiana}
  \country{USA}
  \postcode{47907}
}

\author{Hyungjun Doh}
\email{hdoh@purdue.edu}
\orcid{0009-0008-3154-1201}
\affiliation{%
  \institution{Purdue University}
  \streetaddress{585 Purdue Mall}
  \city{West Lafayette}
  \state{Indiana}
  \country{USA}
  \postcode{47907}
}

\author{Hyung-gun Chi}
\email{stnoah1@gmail.com}
\orcid{0000-0001-5454-3404}
\affiliation{%
  \institution{Purdue University}
  \streetaddress{585 Purdue Mall}
  \city{West Lafayette}
  \state{Indiana}
  \country{USA}
  \postcode{47907}
}

\author{Alexander J. Quinn}
\email{alexander.j.quinn@gmail.com}
\orcid{0009-0000-7964-536X}
\affiliation{%
  \institution{Purdue University}
  \streetaddress{585 Purdue Mall}
  \city{West Lafayette}
  \state{Indiana}
  \country{USA}
  \postcode{47907}
}

\author{Karthik Ramani}
\email{ramani@purdue.edu}
\orcid{0000-0001-8639-5135}
\affiliation{%
  \institution{Purdue University}
  \streetaddress{585 Purdue Mall}
  \city{West Lafayette}
  \state{Indiana}
  \country{USA}
  \postcode{47907}
}

\renewcommand{\shortauthors}{Shi, Jain, Chi, et al.}

\begin{abstract} 
    Context-aware AR instruction enables adaptive and in-situ learning experiences.
    However, hardware limitations and expertise requirements constrain the creation of such instructions.
    With recent developments in Generative Artificial Intelligence (Gen-AI), current research tries to tackle these constraints by deploying AI-generated content (AIGC) in AR applications.
    However, our preliminary study with six AR practitioners revealed that the current AIGC lacks contextual information to adapt to varying application scenarios and is therefore limited in authoring.
    To utilize the strong generative power of GenAI to ease the authoring of AR instruction while capturing the context, we developed CARING-AI, an AR system to author context-aware humanoid-avatar-based instructions with GenAI.    
    By navigating in the environment, users naturally provide contextual information to generate humanoid-avatar animation as AR instructions that blend in the context spatially and temporally.
    We showcased three application scenarios of CARING-AI: Asynchronous Instructions, Remote Instructions, and Ad Hoc Instructions based on a design space of AIGC in AR Instructions.
    With two user studies (N=12), we assessed the system usability of CARING-AI and demonstrated the easiness and effectiveness of authoring with Gen-AI.
\end{abstract}




\begin{teaserfigure}
  \centering
  \includegraphics[width=.85\textwidth]{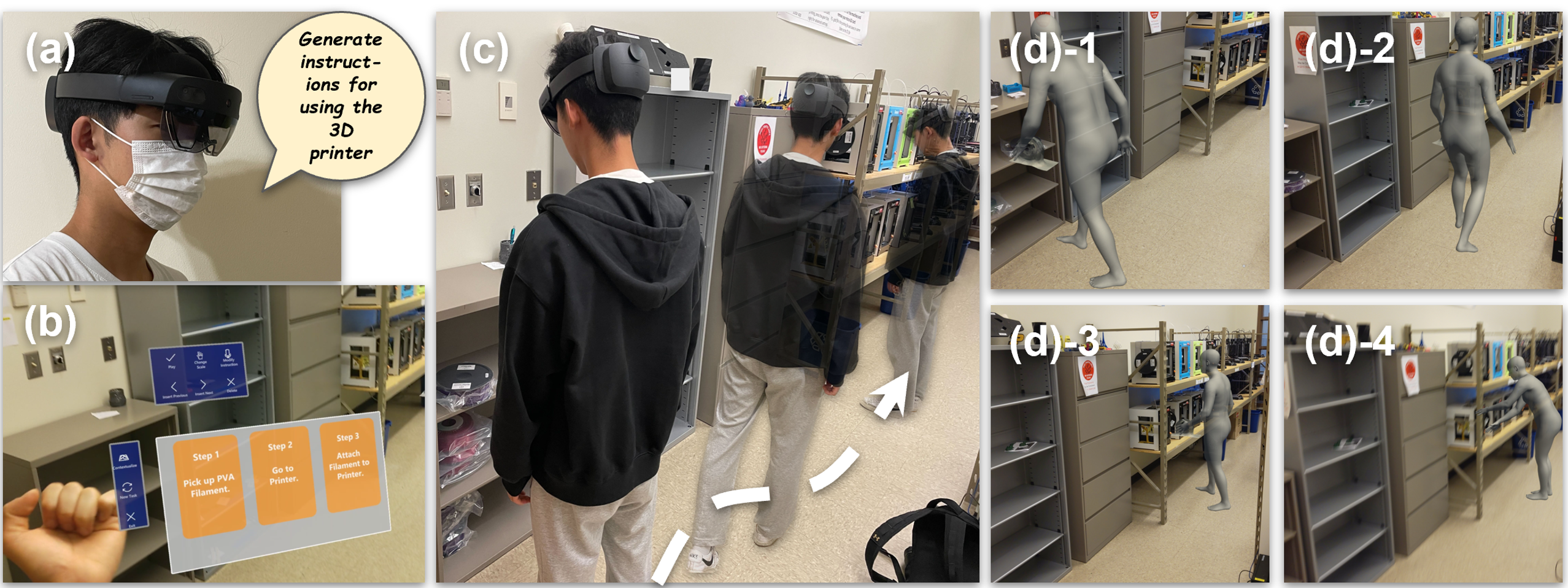}
  \caption{An overview of CARING-AI system authoring workflow. CARING-AI enables authors to create contextualized AR instructions through generative AI. (a) Using CARING-AI, authors first speak their intended instruction content, (b) then the corresponding step-by-step instructions are generated in text. Authors interact with the interface to modify the textual instructions and group them. (c) Then the authors provide contextual information to the instructions by walking in the environment and taking screenshots with the AR HMD. (d) Finally, CARING-AI generates step-by-step humanoid avatar demonstrations of the AR instruction situated in the context.}
  \label{fig:teaser}
\end{teaserfigure}

\maketitle
\section{Introduction}
Augmented Reality (AR) instructions provide an interactive and immersive learning experience by rendering digital content onto physical environments and enabling visualization of complex concepts or procedures.
With such instructions, end-users explore various scenarios and practice skills in a more realistic and context-rich setting.
Due to their vast capabilities and their potential to enhance user engagement~\cite{weerasinghe2022arigato}, facilitate learning~\cite{zhu2023learniotvr}, and improve performance~\cite{hoover2020measuring, christopoulos2022effects} in various contexts, AR instructions have gained considerable attention in a range of fields.


In the manual task instruction domains, humanoid avatars are the preferred options of visualization~\cite{cao2020exploratory, chidambaram2021processar}, because they can convey spatial and temporal instructions on complex sequences of tasks, such as machine tasks, assembly tasks, manual skill learning, and medical training.
Prior works thrived to optimize the authoring of animated humanoid avatars in AR.
Beyond regular animation workflows supported by software such as Unity~\cite{unity3d}, Unreal Engine~\cite{unreal_engine}, or Blender~\cite{blender}, research has proposed diverse methodologies to overcome the requirement of expertise in both the subject matter of the instructions and the programming for animation~\cite{chidambaram2021processar}.
A promising method is Authoring/Programming by embodied Demonstration (PbD, i.e. creating or editing humanoid animation in AR environments by physically interacting or demonstrating actions in the real world).
PbD have the advantages such as realistic animation~\cite{weidner2023systematic, kim2023avatar}, code-less efficiency~\cite{9994974}, engagement~\cite{alblehai2022individual, morris2020toward, black2017can}, interactivity~\cite{wang2020avatarmeeting, jo2015spacetime}, and learning gain~\cite{zhu2023learniotvr} in AR instruction applications.
Despite the benefits and simplicity for the authors, PbD is still subject to real-world human motion (i.e. the authors have to physically present and demonstrate) and requires complex hardware setups and re-setups for Motion Capture (MoCap) such as cameras or motion sensors.
Therefore, authoring with PbD systems is limited in varying contexts ad hoc. 

The development of Generative Artificial Intelligence (Gen-AI) has brought AI-generated content (AIGC) into the discussion of authoring AR instructions~\cite{hu2023exploring}, considering its potential to eliminate expertise barriers and hardware requirements.
With this rapid growth of Gen-AI power, content creation in various modalities can be democratized to higher levels ~\cite{LV2023208, cao2023comprehensive}.
Users are enabled to generate desired content by simply prompting via intuitive modalities (e.g. textual conversation~\cite{radford2018improving, radford2019language, keskar2019ctrl, brown2020language} and reference image~\cite{radford2021learning, ramesh2021zeroshot, ramesh2022hierarchical}).
Many ongoing research and discussions have identified opportunities for deploying AIGC in AR for its power of abstracting human knowledge and a wide range of I/O modalities~\cite{cao2023comprehensive, shi2023hci}.


In pursuit of the design space of AIGC in AR instructions, research is faced with the challenge that Gen-AI lacks the contextual and background information to be deployed into real-world applications~\cite{LV2023208}.
In the scope of AR instruction, contextual information is a critical metaphor, where spatial-temporal information of the instruction is to be blended in the context of the users.
A taxonomy of context-awareness in AR instruction, that many prior works~\cite{qian2023explore, grubert2016towards, wang2022supporting} converge towards, encompasses three key aspects: the human, environment, and system.

Building on this existing knowledge, we aim to fill the gap between state-of-the-art Gen-AI and context-aware AR humanoid avatar instructions.
Specifically, our research is motivated to explore (1) What context information does AI-generated humanoid avatar animation lack for AR instructions? (\autoref{sec:designrationale}) (2) How can this missing contextual information be delivered to Gen-AI? (\autoref{sec:system}) and (3) What insights can we gain from our designs to further foster developments towards the use of AIGC in AR? (\autoref{sec:limitation})

From a preliminary expert interview, we summarize the design goals for naturally providing contextual information to AIGC incorporating user interactions in the authoring process.
We then present CARING-AI, an AR system enabling authoring contextualized humanoid avatar animation for AR instructions.
Given a textual description of the task to instruct, CARING-AI generates step-by-step textual instructions that can be modified by the users and further generates motion that animates humanoid avatars as the visual cues in the instructions.
After giving the textual instructions to animate, authors navigate and scan the environment with an AR Head-Mounted Device (HMD).
Then, CARING-AI temporally and spatially adapts the AI-generated instructions to the human, environment, and system context of the task.
 
Our contributions are four-fold:
\begin{itemize}
    \item A code-less and Mocap-free workflow for authoring animated humanoid avatar instructions in AR with Gen-AI, contextually aware of the human, environment, and system.
    \item A diffusion-model-based algorithm to temporally smooth sequences of individually generated humanoid motions.
    \item An AR interface for authoring AR instructions from textual input describing the tasks, avatars' trajectory, and FOV.
    \item A series of studies evaluating the performance of our system and assessing the efficiency of creating AR animation with Gen-AI compared with a baseline PbD method.
\end{itemize}
\section{Related Work}

\subsection{AR Instruction}
AR instruction refers to the use of AR technology for instructional purposes, such as visualizing complex concepts, exploring various scenarios, practicing skills, and providing real-time feedback.

Our use of the AR instruction metaphor is grounded in real-world applications in diverse domains including assembly~\cite{lavric2022methodologies, frizziero2023augmented, chidambaram2021processar}, education~\cite{uriarte2023comparison, gallardo2022estelar, jeong2023table2table, monteiro2023teachable, pan2021knowing, liu2022designing}, manufacturing~\cite{preda2023augmented}, logistics~\cite{maio2023pervasive}, IoT~\cite{scargill2022environmental, ye2022progesar} and domestic applications~\cite{cao2021context, kang2023arbility, wang2020enhancing, gutierrez2018phara, wang2020enhancing}.

Our scope focuses on the visualization techniques of animated humanoid avatars in authoring AR content for tasks that convey spatio-temporal instructions to the end-users.
Through our wide literature review of AR instructions, we conclude that the information conveyed by AR instructions can be categorized into three types:

\textbf{\textit{Spatial information}} refers to the geographical or spatially-related data such as the location of certain objects or the occurrence of interactions. Spatial information is usually visualized by 3D models~\cite{gallardo2022estelar, herskovitz2022xspace, ye2023proobjar, jain2023ubi}, overlaying data~\cite{kang2023arbility, gutierrez2018phara}, and visual cues such as arrows and lines~\cite{seeliger2022context, cao2021context, lavric2022methodologies}.

\textbf{\textit{Temporal information}} refers to the time-related data such as the order, synchronization, or timing of the movement or occurrence in the AR.
Temporal information can be visualized through textual descriptions of order or procedural~\cite{gattullo2020towards}, animation~\cite{frizziero2023augmented, pan2021knowing}, video~\cite{cao2022mobiletutar}, or sequential overlays~\cite{seeliger2022context}.

\textbf{\textit{Spatio-temporal information}} refers to the information that encompasses both spatial and temporal descriptions of an event, an interaction, or movement in AR, explicitly addressing the change of spatial data in a temporal interval.
    Spatio-temporal information can be visualized in AR by combining spatial and temporal methodologies.
    When spatio-temporal information depicts a human motion or their interaction with the environment, it is better visualized in the animated humanoid avatars~\cite{wang2020capturar, huang2021adaptutar, chidambaram2021processar, 10.1145/3332165.3347902}, where the end-users of the content can learn through following the avatars.

\subsection{Authoring AR Content}
Authoring AR content refers to the process where designers explicitly assign spatial behaviors of the virtual components to the physical world~\cite{qian2022scalar}.
Programming-based authoring tools enable authors to create AR content through programming languages and mathematical modeling~\cite{unity3d, unreal_engine, blender}. 
Authoring by programming creates a precise AR experience, however, at the cost of requiring authors' expertise in both the subject matter and programming.
Moreover, it isolates the authors from the target environment where the AR applications emerge, depriving the spatio-temporal connection to the target environment of the authors.

To tackle the challenges above, prior arts propose the concept of immersive authoring, where the author can create AR content by interacting with both the virtual components and the physical world~\cite{immersiveauthoring}.
To immersively author humanoid avatar animation, prior work has applied methods based on embodied demonstration to authoring.
Through embodied methods, designers can create human movement and interactions with objects by simply demonstrating~\cite{liu2023instrumentar, wang2020capturar, 10.1145/3332165.3347902, 9994974, chidambaram2021processar, reddy2022exploration, wang2022supporting}.
However, authoring through demonstration is subject to the hardware needed for Mocap~\cite{wang2020capturar, chidambaram2021processar}.
In addition, it requires the author to be physically interacting with the environment, which is often not possible or even needed.
For example, the environment may be remote for the author, the environment itself is virtual, the concept that is being demonstrated is not physically plausible or imaginary, or costly for various reasons.

To overcome the barriers of expertise requirement, hardware limitation, and physical interactions, researchers have investigated the uses of AI-generated content (AIGC) in AR applications.
Early works are limited by the modalities and generating power of Gen-AI and, therefore, focus on only a bounded area.
For example, Generative Adversarial Networks (GAN) are capable of generating images based on a given text or image input.
It has been deployed in visual tasks such as fashion design~\cite{sandamini2022augmented,8613637}, rendering a realistic shadow~\cite{liu2020arshadowgan}, reconstructing an occluded human body~\cite{chi2023pose} or virtual objects~\cite{yun2018occluded} or generating new virtual objects~\cite{Soberl2023, kim2021object}.

With the recent development in Gen-AI technology, methodologies have enabled content generation in a wider range of modalities (e.g. text-to-text by Models such as Generative Pre-trained Transformer (GPT) and its successors~\cite{radford2018improving, radford2019language, keskar2019ctrl, brown2020language},  T5~\cite{raffel2020exploring}, and BERT~\cite{devlin2018bert}, text-to-image by large vision models~\cite{radford2021learning, saharia2022photorealistic, ramesh2021zeroshot, ramesh2022hierarchical, midjourneywebsite} and by Diffusion Models~\cite{sohldickstein2015deep, ho2020denoising, Rombach_2022_CVPR, NEURIPS2022_ec795aea, nichol2022glide}, text-to-3D~\cite{liu20233dall}, image-to-text~\cite{radford2021learning}, etc.) with faster and better-generated quality~\cite{NEURIPS2021_49ad23d1}.

The uniqueness of Gen-AI arises from the fact that it can \textbf{generate novel content}, rather than inferencing and acting on existing data or knowledge bases and \textbf{choosing existing content via an if-else rule database}~\cite{gozalo2023chatgpt}.

The recent developments that have demonstrated the out-of-ordinary capabilities of Gen-AI have inspired and enabled our work to embed AIGC into AR applications.
We present related ongoing research (i.e. non-peer-reviewed reports) as well as some recently published papers to differentiate the key aspects of our approach.
To the best of our knowledge, the capabilities we have demonstrated in AIGC for AR are new and are to be still explored from both the design space and applications viewpoints. 
Hu et al.~\cite{hu2023exploring} explored the design space of AIGC + AR applications through an interview, and concluded with several discussions regarding the user, environment, and function of the AR application. 
Lv et al.~\cite{LV2023208} concluded that context is a key consideration in giving prompts to Large Language Models (LLM).
Soliman et al.~\cite{soliman2023unveiling} envisioned using Gen-AI in ARGC for its wide range of modalities.
Chen et al.~\cite{chen2023papertoplace} implemented an LLM-based AR system that incorporates spatial and contextualized information to generate textual instruction in the AR application.
However, these prior works deal with textual instructions, while ours focuses on humanoid animation to provide spatio-temporal instructions with the avatar.
A recent survey by Chamola et al.~\cite{chamola2023beyond} investigated the capabilities of existing Gen-AI methodologies and summarized the characteristics of possible AIGC + Metaverse applications via clustering the methodologies.
Their research pointed out a key insight towards the prospect of Gen-AI in Metaverse: generating 3D content for Metaverse applications (AR in our scope) via Gen-AI needs the incorporation of contextual information.
This insight is also aligned with the recent works such as those of Huang et al.~\cite{huang2023ark} and Shi et al.~\cite{shi2023hci}. 
They recognized the missing "contextual memory" and designed a knowledge interactive agent to identify the missing knowledge and pass it to the Gen-AI model to ground the model in contextual applications.

Motivated by the prior works, we position our work to fill the gap between the AI-generated humanoid avatar animation and AR instructional applications, by contextualizing the generated content via author interactions. 

\subsection{Context-aware AR Applications}
The metaphor of \textit{context-awareness} has been a significant area of interest among both researchers and practitioners.
Lee et al.~\cite{lee2023design} define context awareness as the ability of a system to \textit{apply the patterns given the constraints imposed by the real world}.
An established taxonomy~\cite{grubert2016towards, qian2023explore} categorizes context-awareness into three types:
\begin{itemize}
\item\textbf{\textit{Human Context}}, where the AR systems recognize the humans (users and non-users), take into consideration their profiles~\cite{huang2021adaptutar, krings2020development}, status~\cite{cao2021context, bowman2022buildar, seeliger2022context, scargill2021context}, or their interactions~\cite{liu2022designing, doughty2021surgeonassist, luo2023context, yin2018synchronous, davari2022validating} and adjust the AR components accordingly.

\item \textbf{\textit{Environmental Context}}, where the AR systems perceive the surroundings of the users and understand the presence and absence of physical objects~\cite{wang2020enhancing, chen2018context, kang2023arbility, gutierrez2018phara, kim2023bubbleu}, temporal primitives~\cite{ruizhi2021context, sachan2022designing}, or digital representations of the scenario~\cite{chen2020context, muff2022framework, scargill2021context, davari2022validating, wang2020capturar, li2023location, qian2022scalar, lacoche2022prototyping, gattullo2020towards}, and adjust their components correspondingly.

\item \textbf{\textit{System Context}}, where the AR systems are aware of their input/output~\cite{krings2020development} or their own states~\cite{yin2018synchronous, seeliger2022context, doughty2021surgeonassist} in the realities, and adapt to these contexts.
\end{itemize}

In the scope of AR instructions, all three categories of context awareness are essential.
With human-context awareness, systems are capable of adapting the instructional content according to the users' performance to maximize the learning gain~\cite{huang2021adaptutar}.
Besides, understanding human motion enables the system actively to decide which steps in the instructions are best to be visualized to the users~\cite{doughty2021surgeonassist}.
The location of the visualization in AR also relies on the human context~\cite{liu2022designing, yin2018synchronous, seeliger2022context, davari2022validating}.
On the other hand, environmental context also plays a key role in AR instructions.
For instance, instructing hand-object interactions in AR requires the overlays of 3D models of the objects to be aligned with the physical world for visual cues~\cite{chidambaram2021processar, gutierrez2018phara}.
The environment also possesses rich semantic information that determines the content of the instruction~\cite{wang2020capturar}.
Moreover, the system context helps to decide the procedures in AR instructions by recognizing the states of the instructions and timely transiting to the subsequent ones~\cite{chidambaram2021processar}.

Grounded on the prior works and the three categories above, we discuss how we can contextualize the AIGC in AR instructions.
\section{Preleminary Study and Design Rationale}
\label{sec:designrationale}
\subsection{Preliminary Study}
\label{sec:prelim}
To better understand AI-generative content (AIGC) for AR instructions, we conducted a study with six participants (P1-P6, four males and two females)  who have prior experience in creating AR applications for procedural instruction.
All Participants were academic researchers from different disciplines: Electrical and Computer Engineering (3), Computer Science (2) and Mechanical Engineering (1). 
The mean age of participants was 29.5 and all of them had at least 4 years of experience in creating AR/VR/MR applications.

\begin{figure}[h]
    \centering
    \includegraphics[width=\linewidth]{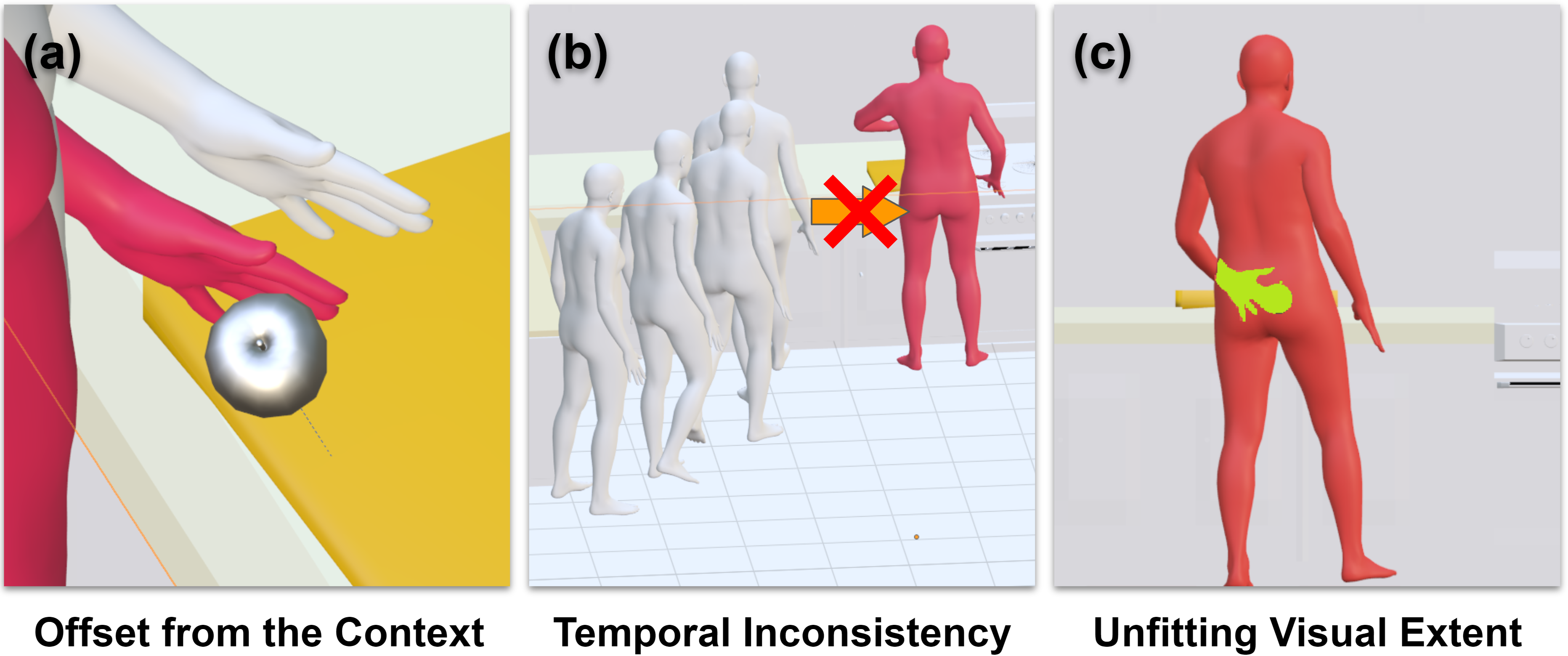}
    \caption{Problems of AI-generated humanoid avatar animation identified in the preliminary study (a) the offset between the generated content and the context, i.e. the interaction is not spatially aligned with the object, (b) the temporal inconsistency, i.e. the generated motion is not temporally connected, and (c) the unfitting visualization extend, i.e. the generated avatars are not of the best scale to convey the instructions (full-body v.s. half-body v.s. hand-only)}
    \label{fig:prelim}
\end{figure}

\textbf{Procedure:} We showed a seven-step humanoid avatar animation instruction task to the participant, generated by the state-of-the-art Generative AI algorithm GMD~\cite{karunratanakul2023gmd}.
The animation is generated from the textual input of \textit{"cutting an apple"} and contains the following steps:
1) Go to the cutting board, 
2) Take the apple with the left hand, 
3) Put the apple on the cutting board, 
4) Go towards the knife area, 
5) Take the knife with the right hand,
6) Go to the cutting board,  
7) Cut the apple with a knife.

After participants watched the content, three authors interviewed them for 30 to 60 minutes with inductive and open-ended questions.
In addition to their opinion on the quality of the shown animation, we asked general questions about the challenges of creating an AR avatar tutorial, the quality of the content, and the potential gap between the characteristics of demonstrations in AR instructions and AIGC. 
The interviews were recorded, transcribed, and coded by the same three authors.
Each author reviewed the transcripts and summarized an initial set of design goals.
Three authors merged to discuss each other's design goals and concluded a refined version by eliminating redundant points and including as many exclusive points as possible.
The analysis provides the following insights and the Design Goals (\textbf{DG}) listed below:

\begin{figure*}[!htp]
    \centering
    \includegraphics[width=.8\textwidth]{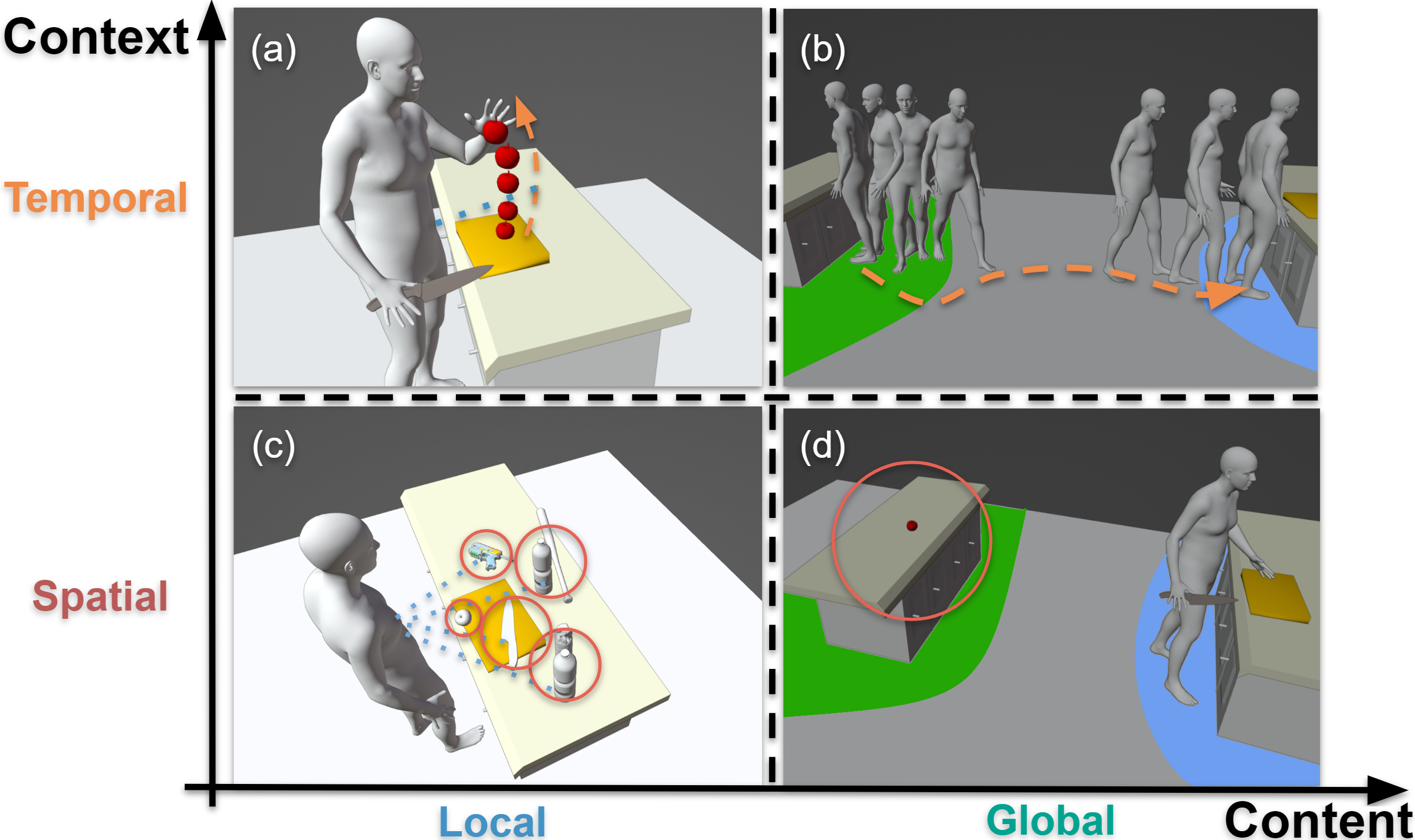}
    \caption{Our consideration of the design space of AIGC in AR instructions is composed of two dimensions: context and content. An AR instruction can be either temporal or spatial based on the contextual information it conveys, either local or global, based on the scale of the content it contains.}
    \label{fig:designspace}
\end{figure*}

\textbf{DG 1) Spatially Aware Content} 
The need for AIGC to be grounded in the real world for AR applications is evident.
The AIGC should be aware of the user's real-world environment which includes objects, their locations, and surfaces.
All participants pointed out that spatial information is important to transfer virtual content into the physical world for AR applications (P1-P6).
Additionally, the AIGC should provide avatar demonstrations subject to the users' vicinity where specific interactions and objects are located (P1, P2, P5). 
"\textit{The tutorial should include an avatar demonstration of manipulating a virtual object, when real and virtual are overlaid for a better understanding of the content.}" - P2  

\textbf{DG 2) Transition Continuity}
The AIGC should be smooth when transitioning from one event or interaction to another.
All the users mentioned that the content shown was not continuous and there were sudden breaks between the interactions.
"\textit{All the actions present were looking separate and there was no connection between the actions}" - P1.
Participants also mentioned that it was difficult to create continuous and smooth AR avatar instructions with the currently available technology (P2, P4). 
"\textit{In my case, I created step-by-step small steps for creating tutorials by avatar demonstrations of assembly}" - P2.    

\textbf{DG 3) Scale of Content} 
The AIGC should include different scales of demonstration adaptive to the different scales of the content in terms of the movement, focusing on different parts of the instructions.
This can be achieved by giving users the freedom to decide whether they prefer to see the whole body (third-person view) or just the hands (First-person view) of an avatar(P1, P5).
Moreover, this will also decide the scale of the avatar and virtual objects present in the scene (P2, P3). 
"\textit{Author should have the freedom to watch the content in the visualization method they preferred.}" - P3.      

\textbf{DG 4) Flexibility in Modifications of the Content}
The AR tutorials should contain flexibility in editing, recreating, or removing the content (P1, P4, P5), which is not enabled by the Gen-AI models themselves without designated interactions with the user.
Participants from their prior experience also mentioned that modification in AR tutorials is time-consuming and requires a lot more effort (P1, P6). 
\textit{"I created an AR avatar tutorial for a mechanical assembly task and it took me a lot of time to make the content"} - P6. 
Participants acknowledge the use of the AI model in creating the tutorials because of less coding effort (P1).
"\textit{It amazes me that these tutorials are just created from the text. This will make AR content creation easy and fast}" - P1   

\subsection{Design Space}
\label{subsec:designspace}
From prior works~\cite{behravan2024generative, jones2023using, hu2023towards, chamola2024beyond} and our study findings, we conclude that the current methods of creating AR instructions from AI-Generated Content (AIGC) are sophisticated and cumbersome.
The four aforementioned design goals are key to grounding AIGC in AR instructions.
Most participants agreed that Gen-AI is a powerful tool that can be used to create AR avatar motions, per intuitive and efficient interaction techniques designated to utilize the generative power (\textbf{DG 4}).
We also found that context and content are the most important aspects for AIGC to be used in creating avatar instructions for AR applications.
From the context side, the Gen-AI model should understand the physical space and their elements which includes recognizing specific locations, objects, landmarks, and their relations (\textbf{DG 1}). 
The content can be either an event or interaction and should be presented temporally consistent to the user (\textbf{DG 1}). 
Moreover, the scale of the content also matters when it comes to the efficiency of the instructions (\textbf{DG 3}).
To this end, we identify context and content as two essential dimensions of the design space of AIGC in AR instructions, as shown in ~\autoref{fig:designspace}.
The first dimension is the context, which can be either spatial or temporal: 
\begin{itemize}
    \item \textbf{Spatial context:} It refers to the information related to the physical environment which involves location, objects, and their interactions.  
    \item \textbf{Temporal context:} It refers to the synchronization and timing of information conveyed by the AIGC.
\end{itemize}    
The second dimension is the content in AR, which can be either global or local: 
\begin{itemize}
    \item \textbf{Local content:} It refers to the specific content of the instruction constrained in the users' immediate vicinity, which is to be depicted in low-level details in the AIGC instruction.
    \item \textbf{Global content:} It refers to the broader perspective of the content relating to the overall scope of the task, describing the high-level goals of steps.
\end{itemize}

\begin{figure*}[!htp]
    \centering
    \includegraphics[width=.9\textwidth]{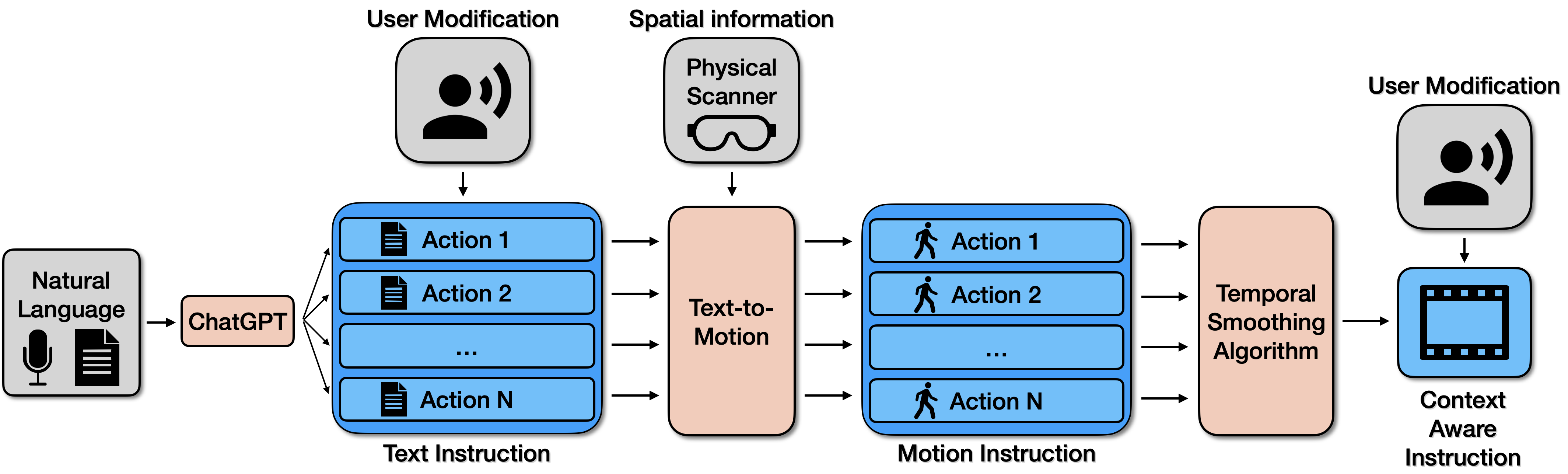}
    \caption{The overall pipeline of the CARING-AI system. Users start by generating textual instructions by speech or text. These instructions will be further grounded in the context of the users by scanning the environment. With context, instructions are used to generate humanoid avatar motion to demonstrate the instructions, blended in AR.}
    \label{fig:pipeline}
\end{figure*}

We further explore the AIGC in AR instructions located in each of the quadrants divided by the two dimensions above.

\textbf{Local-spatial} instructions explain users' closest vicinity information about the objects, locations, their semantic information, and relation with each other \textbf{(DG 1)}.
Such instructions locate and align the 3D object models and humanoid avatars with the corresponding physical objects or areas.  

\textbf{Local-temporal} instructions reveal the timely order of interactions between the avatars and the vicinity.
Such instructions illustrate step-by-step how-to for each interaction or action with temporal consistent transitioning from one to another \textbf{(DG 2)}. 

\textbf{Global-spatial:} instructions depict the approximate whereabouts of the objects, areas, or interactions that are positioned outside the local vicinity.
In contrast to local-spatial instructions, global-spatial instructions posit the content approximately in a space rather than detailing the exact location in the space \textbf{(DG1)}. 

\textbf{Global-temporal:} instructions guide the end-users from one space into another and change the vicinity of the end-users with temporally consistent transitions \textbf{(DG2)}.

We built the CARING-AI system based on the design space decomposition above, addressing the design goals that we have derived.

\section{CARING-AI System}
\label{sec:system}
We developed the CARING-AI system that allows authors to generate and contextualize avatar animation instructions in AR.
Based on the discussion above, we derived the following features in our system: 1) Allowing authors to create textual instruction with editable features (\textbf{DG 4}), 2) Scanning the environment to get spatial context information (\textbf{DG 1}),  and  3) An authoring interface for visualization and editing of the generated content (\textbf{DG 2, 3, 4}).
In this section, we discuss the implementations of the algorithms and modules of CARING-AI and the present our interface.

\subsection{System Overview}

CARING-AI consists of the following steps as shown in ~\autoref{fig:pipeline}: 

\textbf{1) Refining textual instructions.} The user provides a task description to ChatGPT~\cite{openai2021chatgpt}, which returns the step-by-step textual instructions to perform the task. 
The user can then further modify or correct the generated textual instructions.

\textbf{2) Scanning the environment.} The user moves in the physical environment to scan the objects, locations, and areas, as well as record their trajectory, which will be used to provide spatial context information to the system.

\textbf{3) Generating avatar instructions.} The system takes refined textual input from Step 1 to generate avatar instructions based on the \textbf{design space} discussed in~\autoref{subsec:designspace}.
The generated instructions are also grounded by the context information provided in Step 2.

\textbf{4) Visualization and Editing.} The user can view and edit the AI-generated instructions.

\begin{figure}[htp]
    \centering
    \includegraphics[width=.75\linewidth]{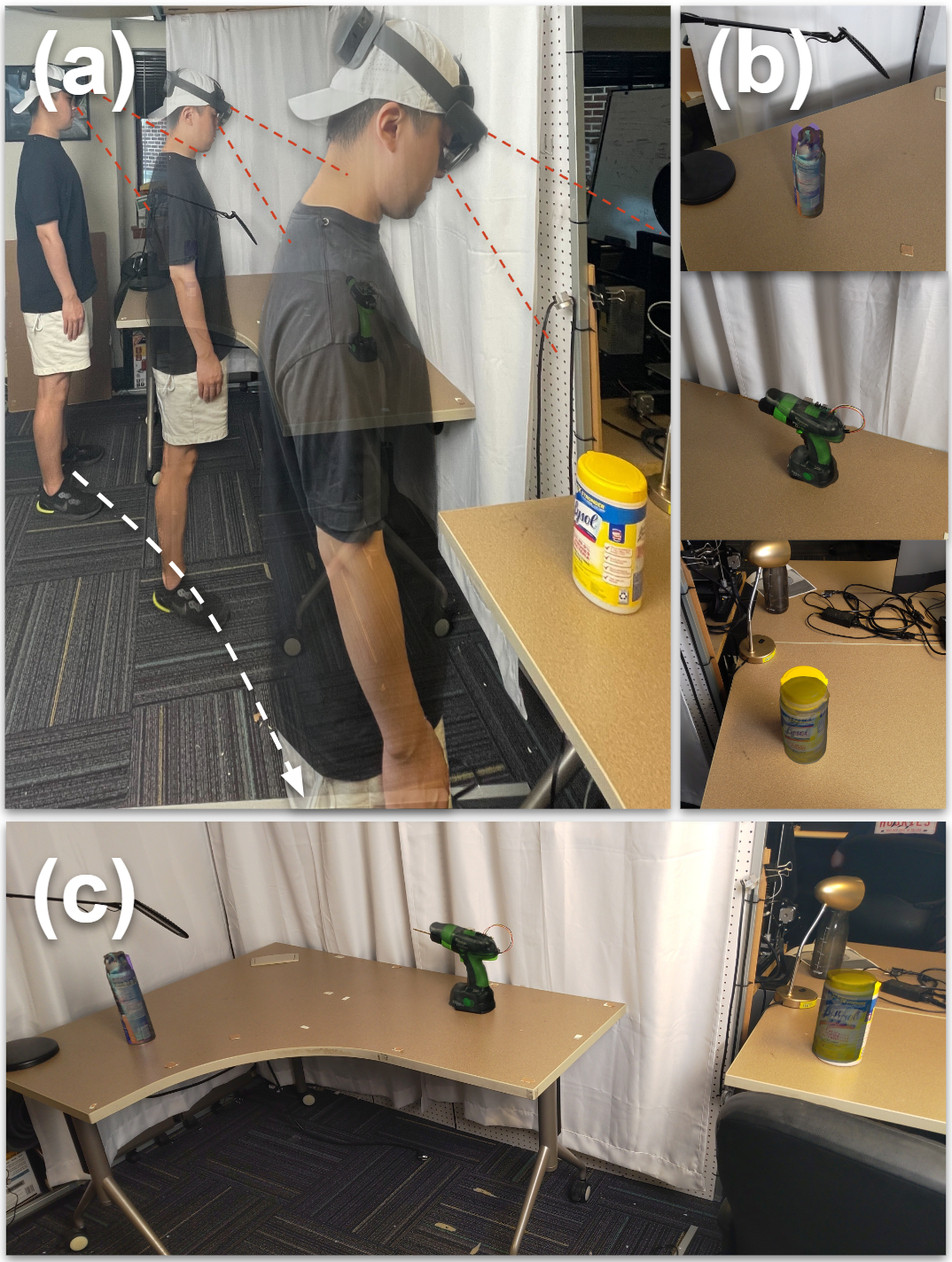}
    \caption{Our methodology for obtaining the contextual information. For global information, users walk from one location to another to provide trajectories (a). For spatial information, users look at the local objects and take screenshots (b, c). This contextual information will be used to generate humanoid avatar motions that are aware of the spatial context for global and local content.}
    \label{fig:scanning}
\end{figure}

\subsection{Textual Instructions}
This module allows users to refine textual instructions for a task using a large language model (LLM), namely ChatGPT API~\cite{openai2021chatgpt}.
Given a user-intended task to instruct, CARING-AI prompts~\cite{white2023prompt} the ChatGPT API to refine the user description of the task into a sequence of step-by-step predefined action labels, which are presented in the HumanML3D dataset~\cite{Guo_2022_CVPR} (a large computer vision benchmark dataset), by specifically asking \textit{"detailed step-by-step instructions of the [task name]"}.
The purpose of this step is to align the terminology of the textual instructions with the available action labels from the dataset to ensure precise generation by the model.

After the refined instructions are generated, users can make necessary adjustments, add more details, or remove information to ensure the instructions align with their specific needs (for example if the object in the textual instruction is not present in the environment).
The finalized instructions will then be used to generate the avatar motion for the task.

\subsection{Scanning the Physical Space}
\label{subsec:scanning}
CARING-AI utilizes HoloLens2 AR-HMD~\cite{hololens2} as the front-end platform.
In order to capture the spatial context of the environment, such as objects, their locations, and semantic meaning as shown in ~\autoref{fig:scanning}, the user navigates and scans the environment with the HMD and starts the \textbf{Scan mode} in the interface.
The user walks around from and to contexts where actions happen, and scans the entire required environment.
Upon entering the \textbf{Scan mode}, CARING-AI records the surroundings by taking RGB images of the HMD FOV and starts recording the global trajectory of the user (built-in SLAM).
The RGB images are passed to an object detection algorithm~\cite{Redmon_2016_CVPR} (30ms per image) to get the semantic classification and relative location of the objects.
The RGB images and the object information are further passed to the state-of-the-art 6 DoF algorithm, MegaPose 6D~\cite{labbe2022megapose} to obtain the 6 DoF information of the objects. Then CARING-AI overlays virtual objects onto the real object based on 6DoF information. 
This information of objects is then used to generate avatar motion with detected and overlayed objects. 

\subsection{Generating the motion}
After getting the textual instruction and spatial information from the user, we generate the avatar motion utilizing a Gen-AI model. 
Specifically, we modified the state-of-the-art text-to-motion AI model (MDM~\cite{tevet2022human}), to generate the global-spatial-context-aware motion (\autoref{subsubsection:spatial}), local-spatial-context-aware motion (\autoref{subsubsection:interaction}), and temporal-context-aware motion (\autoref{subsubsection:temporal}), covering our design space of AR avatar instructions shown in ~\autoref{fig:designspace}.

\begin{figure}[ht]
    \centering
    \includegraphics[width=\linewidth]{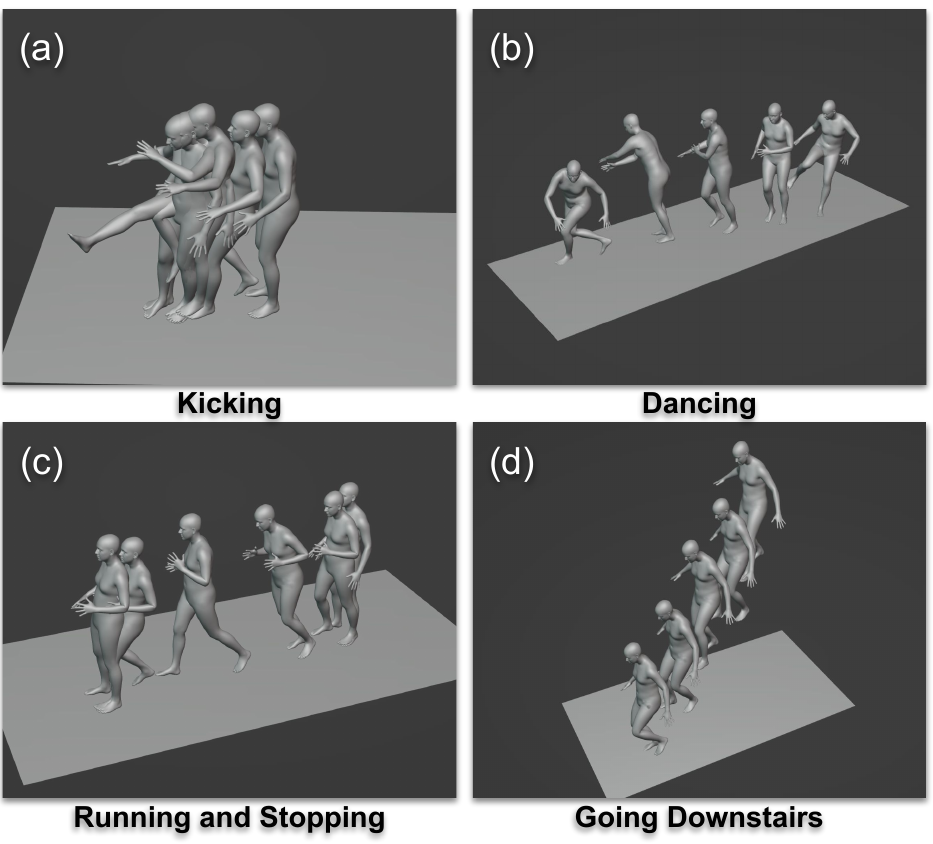}
    \caption{Some examples of our motion generation models. The motion can be local (a) or global (b, c, d, i.e. from one place to another)}
    \label{fig:placeholder_5}
\end{figure}

\subsubsection{Global-Spatial-Context-Aware Generation}
\label{subsubsection:spatial}
\begin{figure}[ht]
    \centering
    \includegraphics[width=\linewidth]{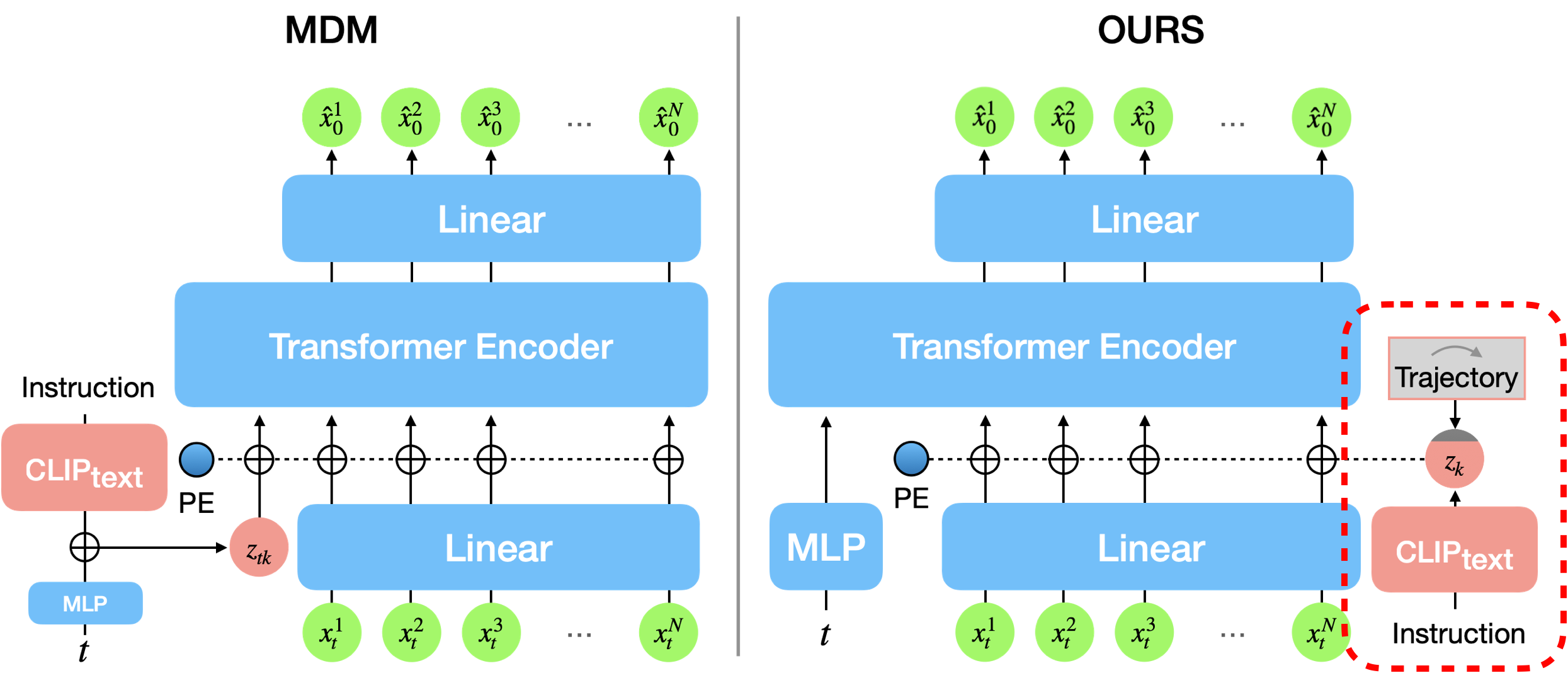}
    \caption{The comparison of the diffusion training overview is between the Motion Diffusion Model (MDM) (\textit{left}) and ours (\textit{right}). The MDM conditions motion frames by placing $z_{tk}$ at the first location, while our conditions motion frames by adding $z_k$ to each motion embedding. For simplicity, we have omitted the random masking of the text embedding used for classifier-free diffusion guidance.}
    \label{fig:architecture}
\end{figure}

As discussed in \textbf{DG 1}, it is key to contextualizing the generated animation for AR instructions.
To tackle this challenge, we exploit the idea of Guided Motion Diffusion (GMD)~\cite{karunratanakul2023gmd}.
On top of other motion generation diffusion models, GMD can generate humanoid motion data, using text descriptions and location cues as the conditions to guide the generation.
However, GMD does not support the generation of sequences of multiple actions.
To address this challenge, we modified the architecture of the Motion Diffusion Model (MDM)~\cite{tevet2022human} (which is also used by GMD as their base model to include trajectories) as shown in ~\autoref{fig:architecture} and applied the GMD method to generate the humanoid motion with trajectory guidance.
We use the trajectories recorded in the \textbf{Scan mode} as the conditions to the diffusion model to provide global spatial information to the generated motion.

\subsubsection{Local-Spatial-Context-Aware Generation}
\label{subsubsection:interaction}
CARING-AI is also capable of conveying local spatial information in the instructions, by generating the motion of the hand with 3D virtual objects overlaid on the physical objects using another motion diffusion model for hand and object interaction ~\cite{cha2024text2hoi}.
As described in ~\autoref{subsec:scanning}, in the \textbf{Scan mode}, CARING-AI obtains the location information of the objects in the physical environment and overlays 3D models of them in AR.
To make sure that the generated avatar interacts with the objects correctly, we ask the users to exit the \textbf{Scan mode} at the end of their trajectory while looking (with the HMD) at the objects that they intend to interact with at the step.
In this way, we guarantee to record the object 6 DoF information relative to the last global location of the trajectory.

\subsubsection{Temporal-Context-Aware Generation} \label{subsubsection:temporal}

As discussed in \textbf{DG 2}, temporal smoothness is key to the sense of continuity in AR instructions.
The original MDM model is designed to generate only a single action by conditioning the instruction into the whole sequence at once.
Generated motions exhibit discontinuity in transition segments because they are produced independently, without incorporating information about the start and end of each instruction as shown in ~\autoref{fig:prelim} (b).

To address this challenge, we modified MDM to condition instructions to each frame, allowing them to generate multiple action sequences jointly.
We visualized the architecture and modification in \autoref{fig:architecture}.
For the sampling process, we generate multiple actions by adding distinct text conditions, represented by \( z_k \), to the frames.
For example, for three actions each 60 frames long, we applied different \( z_k \) values across the ranges: 1--60, 61--120, and 121--180 frames.

\begin{figure}[t]
\centering
\begin{minipage}{\linewidth}
\hsize=\textwidth 
\begin{algorithm}[H]
\caption{Temporal smoothing}\label{alg:smoothing}
\textbf{INPUT: } $K^1_t, K^2_t$ \Comment Transition segments\\
\hspace*{3.8em} $\alpha_t$ \Comment Temporal smoothing function\\
\textbf{OUTPUT: } $\hat{K}$ \Comment New transition segments\\

\begin{algorithmic}[1]
\For{$t = 0,1,...,L-1$}\Comment Temporal smoothing
    \State $\Tilde{K_t} = \alpha_tK_t + (1-\alpha_t)K_{t}$
\EndFor
\For{$t= 0,1,..., 2L-1$}\Comment Linear interpolation
    \State $x \gets \frac{L-1}{2L-1}t, x_0 \gets \lfloor \frac{L-1}{2L-1}t \rfloor, x_1 \gets \lceil \frac{L-1}{2L-1}t \rceil$ \\
    \State $\hat{K}_t=\tilde{K}_{x_{0}}+{\frac {\tilde{K}_{x_{1}}-\tilde{K}_{x_{0}}}{x_{1}-x_{0}}}(x-x_{0})$\\
\EndFor
\end{algorithmic}
\end{algorithm}
\end{minipage}
\begin{minipage}{\linewidth}
\centering
\vspace{3.4em}
\includegraphics[width=.85\linewidth]{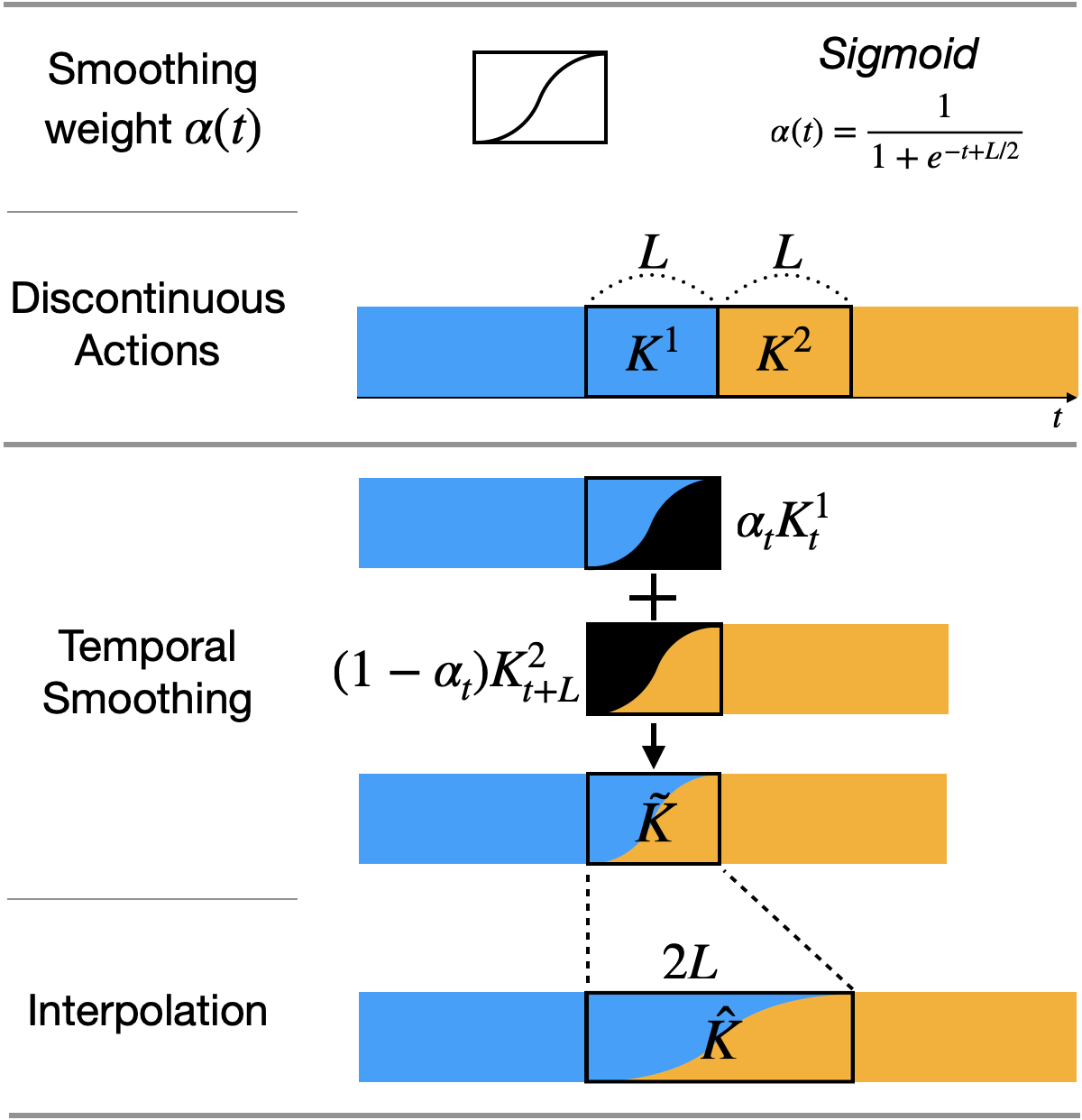}
\vspace{-0.5em}
\caption{The illustration of the temporal smoothing algorithm of CARING-AI}
\vspace{-1em}
\label{fig:smoothing}
\end{minipage}
\end{figure}

However, due to the limitation of the frame length of the training dataset, the quality of the motion drops empirically when the frame number exceeds 196.
Further, we designed a temporal smoothing algorithm to generate an unlimited length of smooth avatar motion and applied it after the generation of motions.
As illustrated in \autoref{fig:smoothing}, the temporal smoothing function, (denoted as $f$) aims to mitigate the discontinuity among the transitional segments of motion ($K^1$ and $K^2$, where $K$ represent two transition segments).
Each of the transition segments comprises a length of $L$ frames.
We also set the weight function $\alpha_t$ to define the ratio for combining the two transition segments.
For this purpose, we employed the shifted sigmoid function for $\alpha_t$, given by $\alpha(t) = \frac{1}{1+e^{-(t-(L/2)}}$, to serve as our smoothing mechanism.
Consequently, the resultant mixed frames, represented as $\tilde{K_t}$, can be expressed as
\begin{flalign}
    \tilde{K_t} = f{(K^1_t,K^2_t,\alpha_t)} = \alpha_tK^1_t + (1-\alpha_t)K^2_{t}.
\end{flalign}
Then, to keep the length of the generation action length, we extended its length twice with linear interpolation sampling.

\begin{flalign}
\hat{K}_t=\tilde{K}_{x_{0}}+{\frac {\tilde{K}_{x_{1}}-\tilde{K}_{x_{0}}}{x_{1}-x_{0}}}(x-x_{0}),
\end{flalign}
where $x$ is $\frac{L-1}{2L-1}t$, $x_0$ is $\lfloor \frac{L-1}{2L-1}t \rfloor$, $x_1$ is $\lceil \frac{L-1}{2L-1}t \rceil$, $\lceil\cdot\rceil$ and $\lfloor\cdot\rfloor$ indicate the ceiling and the floor operator, respectively.




%

\subsection{AR Interface}

To achieve \textbf{DG 3} and \textbf{DG 4}, We introduce an AR interface that includes all the functions discussed above and additional functions such as visualization, editing, and modifying the content.
The authoring system for CARING-AI consists of four modes:
1) \textbf{Task mode} to get users the step-by-step instructions,
2) \textbf{Scan mode} to ground the instructions in the context,
3) \textbf{Author mode} to design and edit textual instruction and avatar motion content,
and 4) \textbf{View mode} to examine the authored AR avatar instructions.
The AR menu is always present in the user's view on the left hand so that users can easily access the all functions of the current mode and also switch between them.

\begin{figure}[htp]
    \centering
    \includegraphics[width=.4\textwidth]{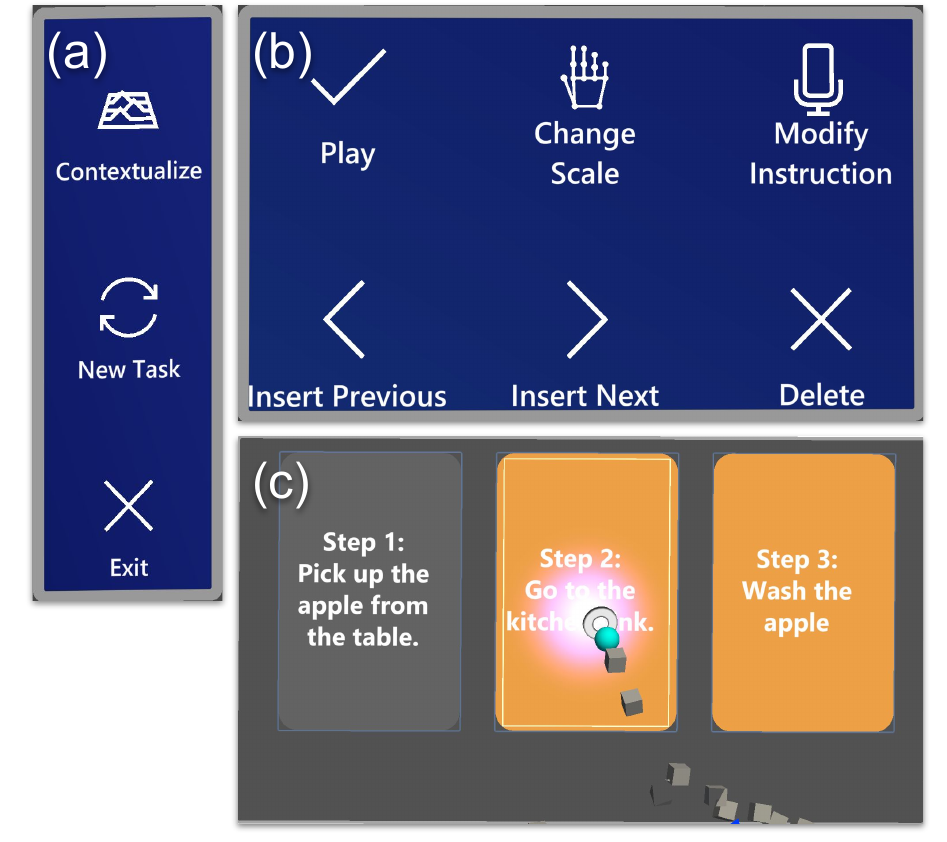}
    \caption{AR User Interface of CARING-AI. In (a), users can start authoring a new task or start contextualizing the instructions. In (c), they can see the generated textual instructions and select a single or multiple of the instructions. In (b), they can choose to view the humanoid avatar animation (\textit{Play} button), change the scale of the humanoid avatar (\textit{Change Scale} button), modify the textual instruction by speech (\textit{Modify Instruction} button), insert new instructions (\textit{Insert Previous} and \textit{Insert Next} buttons), and delete the selected instructions (\textit{Delete} button).}
    \label{fig:ar_interface}
\end{figure}

As shown in ~\autoref{fig:ar_interface} (a), the user first starts by providing the task description using a voice command by clicking the \textit{New Task} button to enter the \textbf{Task mode}.
The user speaks to the system to specify their task, then the system generates textual instructions shown in the instruction panel ~\autoref{fig:ar_interface} (c).
When the users select one step from the panel, they can insert new instruction steps, delete the selected ones, or modify them.

Then, the user selects and groups several steps that happen in a global context (i.e., steps that happen at the same location in the space, for example, in ~\autoref{fig:ar_interface} (c), \textit{Step2: go to the kitchen sink} and \textit{Step3: wash the apple} belong to the same global context), with the selected instructions highlighted in yellow.
After selection, the user clicks the \textit{Contextualize} button and enters \textbf{Scan mode} to scan the physical environment. In the \textbf{Scan mode}, the user simply walks in their physical space to mark the global location for the current group (e.g., in~\autoref{fig:ar_interface}, the user walks to the kitchen sink) and ends contextualizing the current group by taking a screenshot while looking at the contextual environment (e.g. looking at the sink with the apple and knife visible in the scenario).
Upon object detection, CARING-AI then overlays 3D virtual objects on the corresponding physical objects which users can see and adjust the 6 DoF with built-in freehand interactions. 
Iteratively, the user groups and contextualizes the rest of the instructions.
The contextualized instruction panels are highlighted in green while the users are still allowed to revisit and edit.

Upon the completion of contextualizing all steps, the user enters the \textbf{Author mode}.
The user can click the \textit{Modify Instruction} button to modify the instruction and regenerate animation for a specific step, or click the \textit{Change Scale} button to change the visualization scale of the selected step.
The available scales of the visualization are full-body avatars and hand-object avatars (i.e. only the hands, the forearms, and the objects are rendered).

Meanwhile, the user can enter \textbf{View mode} by clicking the \textit{Play} button.
This mode visualizes the currently selected instruction by rendering the generated context-aware avatar animation in the HMD.

\subsection{Software and Hardware Implementation}
\label{sec:implementation}
We implement CARING-AI using Hololens 2~\cite{hololens2} with built-in SLAM tracking for AR experiences.
CARING-AI interface was developed in Unity 3D on a local PC (Intel core i7-9700K CPU, 26 GHz, 128 GB RAM). During the scanning mode, we use a resolution of 1280 x 720 for the RGB image.
The images are then processed in a local PC for object detection and the 6 DoF estimation algorithm for overlaying the virtual 3D on the real object.
We used the Mixed Reality toolkit (MRTK) for the interactions of hands with the virtual objects and the interface. 
For 6 DoF of the object, we used the pre-trained MegaPose6D~\cite{labbe2022megapose} model, which can estimate 6 DoF of objects in the wild.
For object detection, we used the detection model~\cite{Redmon_2016_CVPR} pre-trained on ImageNet~\cite{deng2009imagenet}.
We fine-tuned the object detection algorithm which is used in finding the spatial context for the content.
The training of object detection was performed on objects dataset collected for used cases and user study purposes.
For each object class, we collected 600 images.
The 3D scans of the objects were also collected and stored in the database for the 6DoF algorithm and virtual object overlays in physical.
As mentioned in section 4.4, we used the pre-trained Guided Diffusion Model~\cite{karunratanakul2023gmd} as the motion generation model on the HumanML3D~\cite{Guo_2022_CVPR} dataset.
The action classes from the dataset are further used in the user study and for the demonstration.
One batch of motion generation takes time of 36 seconds, with one NVIDIA RTX A6000 GPU.

\section{Quantitative Evaluation}
\label{sec:quaneva}
In this section, we assess the efficacy of our context-aware generative AI approach in real-life scenarios by comparing it with a baseline (GMD~\cite{karunratanakul2023gmd}). 
As a preliminary step, we evaluated our modified diffusion model algorithm~\autoref{fig:architecture} compared to GMD quantitatively.
We chose GMD as our baseline for comparison because GMD is a state-of-the-art model based on MDM.
This study assesses the modified model's performance in generating humanoid animation, which is the backend algorithm of our system.

\begin{table*}
\caption{Task and instructions}
\resizebox{1.0\linewidth}{!}{
    \label{table:instruction}
    \begin{tabular}{l | l} 
        Task & Instructions \\
        \hline
        \textbf{Charging a Phone} & Get the charger; Insert the cable into the phone; Plug the charger into an outlet\\
        \textbf{Turning on the TV} & Pick up the remote; Point it at the TV; Press the power button\\
        \textbf{Closing a Window} & Approach the window; Grasp the handle or sash; Push to close\\
        \textbf{Starting a Computer} & Sit in front of the computer; Press the power button; Wait for it to boot up.\\
        \textbf{Exercising} & Crawl; Run; Band Push; Crawl to Stand \\
        \textbf{Reading a Book} & Walk to the bookshelf; Choose a book; Go to the living room; Sit on the couch or chair;\\
        \textbf{Closing a Window} & Approach the window; Grasp the handle or sash; Push or slide to close\\
        \textbf{Eating an apple} & Approach to the table; Pick up the remote; Eat the apple; Move back; Turn around; Leave the kitchen\\
        \textbf{Use a 3D printer} & Pick up PVA; Go to printer; Attach Filament to printer; Start printer\\
        \textbf{Making Tea} & Boil the water; Place a cup on the table; Pick the pot; Pour boiling water into the cup.\\
        \hline
    \end{tabular}
}
\end{table*}
\subsection{Evaluation}
\subsubsection{Baseline}
We used a pre-trained model of GMD to compare our algorithm. 
GMD \cite{karunratanakul2023gmd} is pre-trained with the HumanML3D dataset, which is annotated human motion data.
The dataset has 22 joints $|\mathcal {J}|=22$ following the skeleton representation of the HumanML3D dataset~\cite{Guo_2022_CVPR}. 
The HumanML3D dataset encompasses 14,616 motions, paired with 44,970 descriptions that are comprised of 5,371 unique words. The combined duration of all motions is 28.59 hours. On average, each motion spans 7.1 seconds, and each description contains 12 words.

\subsubsection{Metrics}
To validate the performance of our model, we constructed 10 practical scenarios \autoref{table:instruction} using both the baseline method \cite{karunratanakul2023gmd} and our context-aware approach.
Our evaluation has been done in two dimensions: spatial and temporal context awareness. 
For assessing temporal context awareness, we quantified the motion discontinuity between consecutive instructions.
A heightened awareness of the temporal context by the AI should result in reduced discontinuities in the generated instructions.
The motion distance across frames was computed following the \cite{athanasiou2022teach}. 
We calculate the transition distance, which calculates the joint distance of two transition frames.
\begin{flalign}
d_{\text{temporal}} = \frac{1}{|\mathcal{K}||\mathcal{J}|}\sum_{K\in\mathcal{K}}\sum_{J\in\mathcal{J}}{||J_{K^{last}}-J_{K^{first}}||_2}, 
\end{flalign}
where $\mathcal{K}$ is the set of the two consecutive indices of transition frames $(K^{last}, K^{first}) \in \mathbb{N}^2$, which is composed of the last frame of the previous action $K^{last}$ and the first of the next action $K^{\text{first}}$. The number of transitions is equal to substituting one from the number of instructions $|\mathcal{K}| = |\mathcal{A}-1|$. $\mathcal{J}$ is the set of joints, containing the 3D location of joints at the transition, $J_{K^*} \in \mathbb{R}^3$ as elements.
The human skeleton data we used has 22 joints $|\mathcal {J}|=22$ following the skeleton representation of the HumanML3D dataset.

In terms of spatial context awareness, we gauged the proximity between the avatar and the object specified in the instruction. 
The absence of spatial context often results in instructions that position the avatar at a considerable distance from the target object, potentially leading to user confusion.
We employed the mean Euclidean distance to measure the spatial alignment within the frames of interest.
\begin{flalign}
d_{\text{spatial}} = \frac{1}{\mathcal{|T|}}\sum_{t\in\mathcal{T}}{||J_t^{xy}-O_{t}||_2},
\end{flalign}
where $J^{xy}_t, O_{t} \in \mathbb{R}^2$ indicates the 2D X, Y coordinates of the root joint and target keypoint at the $t$-th frame, respectively.
$\mathcal{T}$ is the set of frames that is spatially conditioned by target keypoint $O_t$.

\subsection{Procedure}
To evaluate our developed algorithm performance, we choose 10 practical scenarios \autoref{table:instruction} commonly found in real-world tasks. These tasks have more than two instructions and are performed at varied locations covering our design space which makes them suitable for evaluating our algorithm and comparing it with the baseline.
To get the instructions for the task, three authors individually provided the task description to ChatGPT and noted the instructions. Then the authors discuss to finalize the steps of the instructions. Additionally, one of them wears hololens to get the spatial context for the algorithms. After generating the text instructions, we input them into a Text-to-Motion generator, resulting in motion instructions.
To evaluate our approach, we compared our motion instructions with those from the GMD\cite{karunratanakul2023gmd}, one of the state-of-the-art algorithms in Text-to-Motion generation.
For a consistent comparison, we kept the length of each instruction the same in 90 frames.

\begin{table}[]
    \centering
    \begin{tabular}{l c c}
        \hline
        Method & Transition distance$\downarrow$ (m) & Absolute distance$\downarrow$ (m)\\
        \hline
        GMD & 0.15 & 0.08\\
        Ours & 0.03 & 0.09\\
        \hline
    \end{tabular}
    \caption{Transition Distance is the comparison of the discontinuity with and without temporal smoothing methods. The lower the better. Absolute distance is the average distance between the avatar and the key points. Distance under 0.1 m is considered as plausible motion \cite{karunratanakul2023gmd}}
    \label{tab:discontinuity}
\end{table}


\subsection{Results and Analysis}
In this section, we detail the results of our preliminary evaluations.
We highlight the transitional gap between two consecutive frames measured in meters ($m$). As illustrated in ~\autoref{tab:discontinuity}, our approach ensures smooth frame transitions. GMD \cite{karunratanakul2023gmd} exhibits a transition distance of $0.15 m$ when frames are simply concatenated. In contrast, our method substantially decreases this transition distance to $0.03 m$ ($p<0.05$), eliminating any motion discontinuity.
Additionally, ~\autoref{tab:discontinuity} showcases the spatial alignment. The distance is determined between the avatar's center and the guided keypoint, assuming an avatar height of $175 cm$. Our method produces results closely aligned with GMD, generating plausible motion with an error margin under $0.1 m$  ($p<0.05$)~\cite{karunratanakul2023gmd}, while also capable of producing smooth and varied actions in one seamless operation.
The \textit{Exercising} task shows the highest spatial error because it contains the instruction to \textit{run}, which represents the most sudden motion among all instructions. Meanwhile, the \textit{Starting a Computer} task has the lowest error due to its fewer movements.
We observed that the quality of hand motion generated by both GMD and our method is subpar. Instructions involving hand-object interactions especially exhibit awkward hand gestures. For instance, the \textit{pickup} motion doesn't adequately display grabbing gestures.
\section{User study 1: Usability}
\label{sec:qualeva}
We conducted a user study to qualitatively evaluate the usability of our system.
We qualitatively evaluated all the steps used in our system as well as the quality of the human motion generation.
We invited 12 users (10 males and 2 females) from the technical university.
All of them have prior experience in AR/VR applications using tablets, AR screens, and head-mounted devices. 
CARING-AI is designed to help both experts and non-experts create human avatar motion. 
The users are from graduate and undergraduate programs and their age ranges from 19 to 30.
None of the users have used our system and have had no knowledge about it before.
The entire study took one hour - two hours and each user was compensated with a 15 USD e-gift card.
The study was conducted in an indoor environment. After the user arrived,  we provided a brief overview of the study. 
Then the users were asked to sign the consent form only when they were comfortable in performing the user study.
After that, we explained the entire CARING-AI system workflow and each function in the UI. Out of 12 users, 5 users had prior experience with developing or using Hololens.  
The users were given enough time to get comfortable with the CARING-AI system before the study was officially started. 
Also, some of the users with no experience were provided with a built-in Hololens tutorial to learn the basics. 
As our study focus was on the usability of the system and user experience on the generated content, we asked the users to complete a System Usability Scale (SUS) and a 5-point scale Likert-type questionnaire followed by 20-minute post-session conversation-type interviews to provide subjective feedback about CARING-AI.

\subsection{Procedure}
\begin{figure}[ht]
    \centering
    \includegraphics[width=\linewidth]{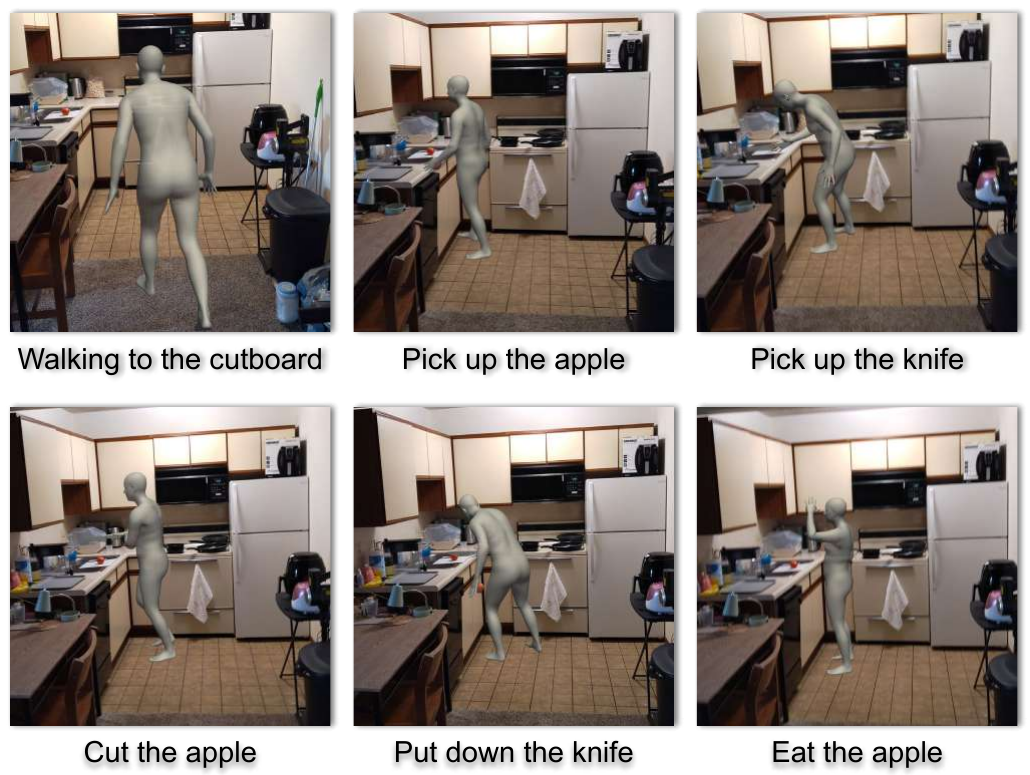}
    \caption{Examples of humanoid animation generated in User Study 1.}
    \label{fig:userexample}
\end{figure}

We evaluated the performance of our system and let users generate avatar motions for the tasks \textbf{Cutting an Apple}. The study took place in a kitchen environment.
The task was chosen because it involves multiple steps and different locations.
The task is suitable for evaluating context-generated content and other system components like interface.
The users were tasked to generate step-by-step instructions for the task from ChatGPT.
The most common steps found in the task as shown in \autoref{fig:userexample} are
\begin{enumerate}
    \item Walking to the cut board (Global, Temporal and Spatial),
    \item Pick up an apple on the table (Local, Temporal and Spatial),
    \item Pick up the knife (Local, Temporal and Spatial),
    \item Cut the apple (Local, Temporal and Spatial),
    \item Put down the knife (Local, Temporal and Spatial),
    \item Eat the apple (Local, Temporal),
    \item Move back (Global, Temporal),
    \item Turn around (Global, Temporal),
    \item and Leave the kitchen (Global, Temporal and Spatial).
\end{enumerate}

Then the user scans the environment and takes the screenshots at different locations.
After that user aligns the virtual objects on the real object if they are not properly aligned by the system. And finally, the user uses the CARING-AI interface to generate the motions.

\begin{figure*}[ht]
    \centering
    \includegraphics[width=.9\textwidth]{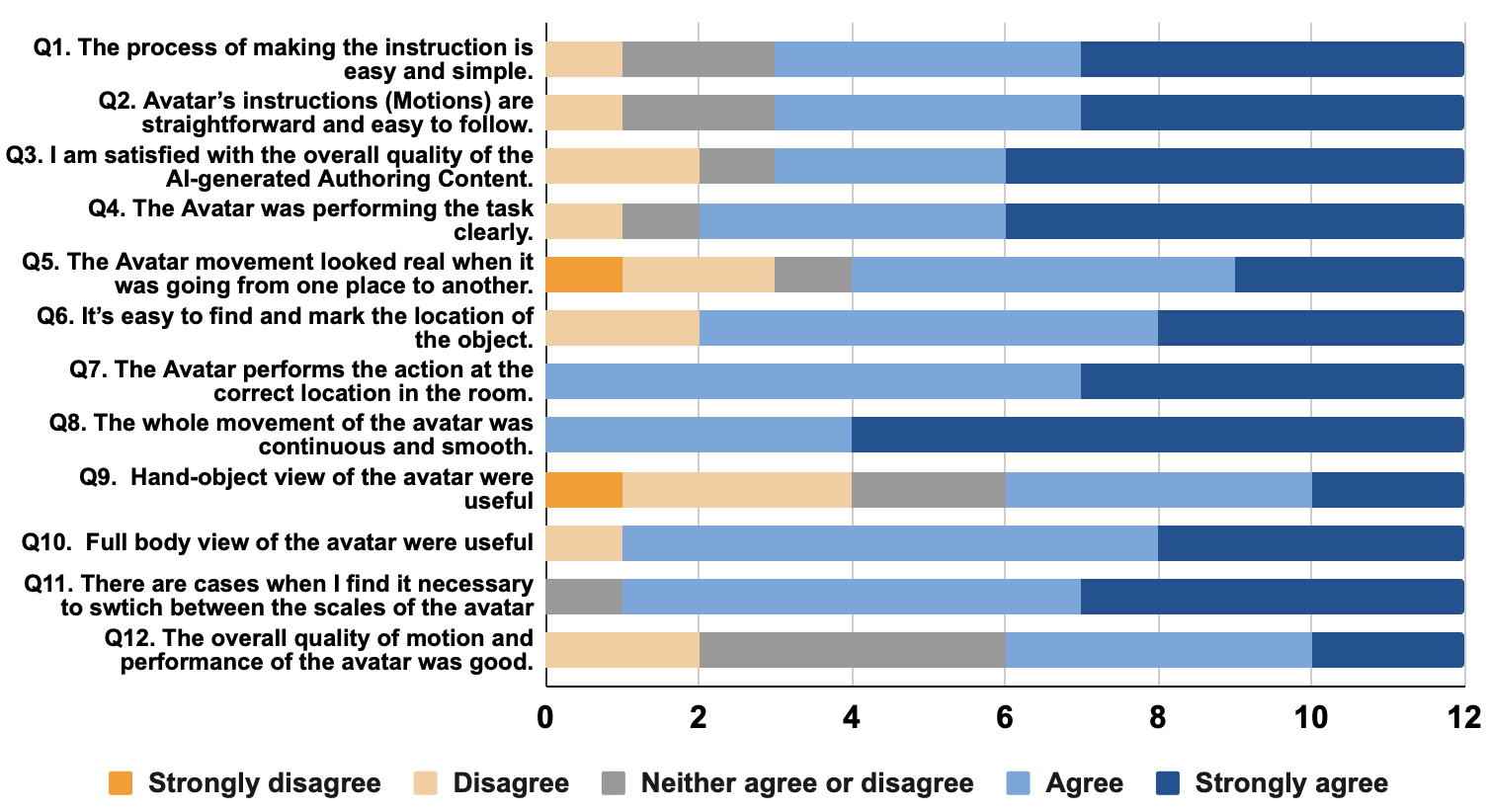}
    \caption{Likert-type questionnaire results from User Study 1.}
    \label{fig:userstudy2}
\end{figure*}

\subsection{Results and Analysis}

We analyzed responses to a 5-point scale Likert-type questionnaire, SUS, and transcribed the interview from the user. 

\subsubsection{Textual Instructions}

We qualitatively evaluate textual step-by-step instruction generated from ChatGPT.
In general, users preferred the step-by-step instructions generated from the ChatGPT to be relevant to the task.
\textit{"P1: I think I don't need to modify the instructions. They were correct and right for the task."}. However, some users modified a few steps little for their instructions.   
Many users acknowledge the visualization of a graph representation of step-by-step instructions and agree that the interactions with the graph are easy to use and simple to follow (Q1: AVG = 4.08, SD = 1.00).
\textit{"P5: The process of creating the instructions was easy and quick."}

\subsubsection{Context Aware Instruction by Avatar}
\label{sec:caeva}
Through post-study interviews and designated Likert-scale questionnaires with the users, we qualitatively evaluate the context in the content generated by CARING-AI during the user study.
Many of the users stated that the avatar was performing the actions with the object at the correct location (Q7: AVG = 4.42, SD = 0.51).
As a piece of evidence, P3 commented in the interview \textit{"P3: I was actually surprised by the way Avatar went to the exact position and performed the activity."}
Another user mentioned \textit{" P2: I liked that I could see the avatar move towards the apple and the fluid and connected motion"}.						
The majority of the users were satisfied by the actions performed by the avatar using the virtual objects (Q4: AVG = 4.25, SD = 0.97).  
As P9 commented in the interview on \textit{"P9: The action demonstrated by the avatar was with the right object."
However, a few users raised concerns about accurate avatar hand and virtual object interactions, such as P12
\textit{"P12: It is not clear to me why the hand was not grabbing the object and it automatically sticks to the hand."}}
We discuss this limitation in more detail in ~\autoref{sec:limitation}.
Users acknowledge that the motion of the avatar from one place to another looks real (Q5: AVG = 3,58, SD = 1.31), such as P11
\textit{"P11: I can't believe that the avatar movement exactly looks as if a real human is walking. I should say this is too cool."}
Users found a smooth transition of the avatar motion between the instructions (Q8: AVG = 4.67, SD = 0.49).
P8 pointed out that the transition by our smoothing algorithm made the animations seamless and the breaks between animations hard to identify, 
\textit{"P8: It was hard for me to draw the boundary between the instructions when I was looking at the avatar motion."}

\subsubsection{Overall System Usability and Utility}
The overall system Likert scale results are shown in ~\autoref{fig:userstudy2}.
Context from the user is the foundation of our generated animation and most of the users were satisfied and comfortable with taking screenshots during the scanning of the environment process (Q6: AVG= 4.00, SD = 1.04), as P8 commented
\textit{"P8: I didn't find any difficulty in moving around and taking screen pictures."}
Further, users also found the alignment of real and virtual objects was accurate.
P4 commented that the accurate alignment contributed to their overall experience \textit{"P4: I think the virtual model was approximately over the top of the real for many objects and the visualization being a 3D rendering definitely helps my experience".}
The CARING-AI system interface was appreciated by the users.
The positive feedback from the users on the usability of the interface is mainly attributed to the easiness of using it, as P2 commented:
\textit{"P2: In my opinion, I find the UI very straightforward and easy to use."}
Moreover, users find it easy to switch between full-body pose and only hand (Q2: AVG= 4.08, SD = 1.00). 
The final avatar motion instruction generated from CARING-AI received a positive response from the user after watching the final generation of instructions (Q12: AVG= 4.42, SD = 0.51). 

Regarding utility, many users reported positive regarding utilizing CARING-AI in creating AR instruction tutorials of human demonstration.
P7 with previous experience of authoring AR instructions in Unity positively commented on the efficiency when utilizing CARING-AI
\textit{"P7: I have developed an AR instruction by coding in Unity and it took me several days to make it. I wish this thing was developed earlier so that I could have used it."}
Some users needed more features to display such as text, and icons for object movement directions along with just demonstrations, such as P9
\textit{"P9: For the base level, it is okay but I think it would have been better if your system provided visual cues showing the movement of the object"}
We discuss the limitation in more detail in ~\autoref{sec:limitation}. 
For the system usability, the users agree that the system is usable (SUS: M = 83.21 out of 100 and SD = 7.34).
A score above 70 is practically considered "Good" usability and an 85-and-beyond score is considered "excellent" as mentioned in~\cite{bangor2008empirical,bangor2009determining}.

\section{User Study 2: Interaction}
To evaluate the interaction design of our system compared wt the baseline programming by Demonstration, we conducted an additional within-subject comparative user study (N=12) between CARING-AI and a baseline PbD method.
The purpose of this study is to assess the novel interaction proposed in CARING-AI and compare the user feedback on the interactions with that from the existing methods (PbD).
To make a reasonable PbD baseline, we followed a similar approach~\cite{reipschlager2022avatar} and built our setup, where the humanoid animation is captured by a third-person-view RGB camera.

\begin{figure*}[ht]
    \centering
    \includegraphics[width=.8\textwidth]{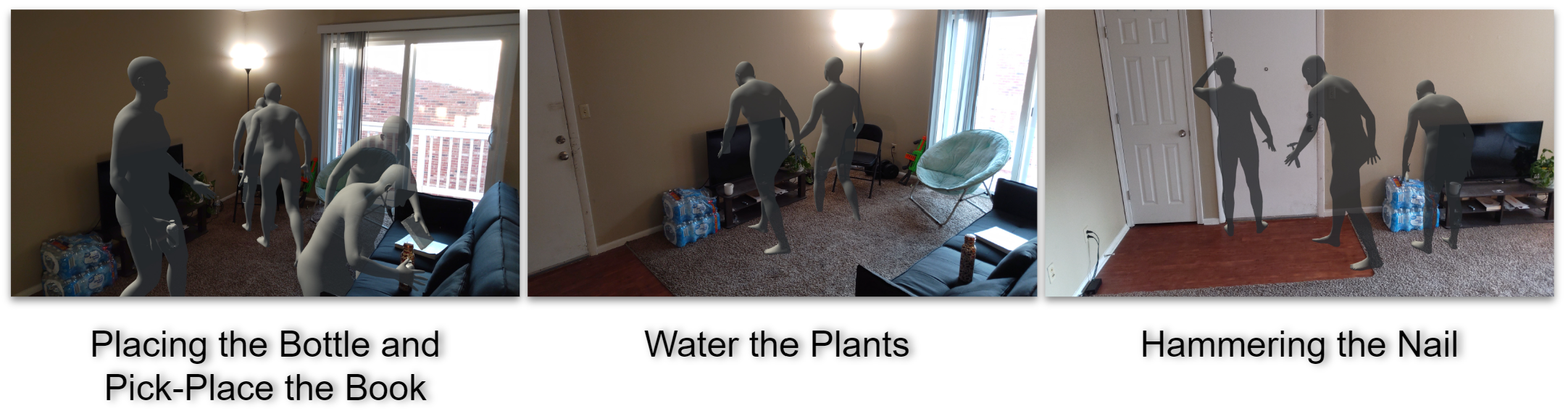}
    \caption{AR animation generated by users using CARING-AI in User Study 2.}
    \label{fig:task_user_study_caring}
\end{figure*}

The participants (8 males and 4 females) are six novices and six experts in developing AR applications.
They were recruited and compensated as in User Study 1.
Users were asked to author AR instructions with both CARING-AI and PbD, counterbalanced by 6 participants authoring with CARING-AI first followed by PbD, and the other 6 participants with PbD and then CARING-AI.
The entire study took 1 to 2 hours.
We followed the same protocol for explaining our system as in User Study 1.
To quantitatively evaluate our system, We asked the users to fill out NASA TLX~\cite{nasa1986task} and a five-point Likert-type questionnaire (\autoref{fig:study2question}).
This questionnaire is designed to collect qualitative evaluations of the users on the efficiency and accuracy of both authoring methods (Q2-5), as well as the quality of the final output (Q1).
Additionally, a 15-minute semi-structured interview was conducted for each participant. In post-processing of the study data, we calculated error rates during interactions and time spent in creating animation.

\subsection{Procedure}

The user study was performed in a living room and users were asked to perform three tasks: organizing the living room, watering a plant, and hammering a nail to a door.
We specifically chose tasks that require human motion that can be guided by humanoid avatar animation in AR.
Also, all chosen tasks involved hand-object interactions with different objects and took place at various global locations, which are:   

\begin{enumerate}
    \item Walk to the sofa. Place the water bottle on the sofa. Pick up the book.\\Walk to the chair. Put down the book on the chair.
    \item Pick up the mug. Walk to the plant. Pour the water into the plant.
    \item Walk to the door. Pick up the hammer. Hammer the nail on the door.
\end{enumerate} 

With CARING-AI, the users first scanned the environment by moving around and taking screenshots of the locations where local actions were to happen.
After that, users aligned the virtual object with their real counterparts.
Then, users generated the final animated instructions following the workflow of CARING-AI as in User Study 1.
Until satisfied, the users could regenerate or adjust the animation with CARING-AI.
The examples of generated AR animation with CARING-AI in this study are shown in ~\autoref{fig:task_user_study_caring}.

\begin{figure*}[ht]
    \centering
    \includegraphics[width=.8\textwidth]{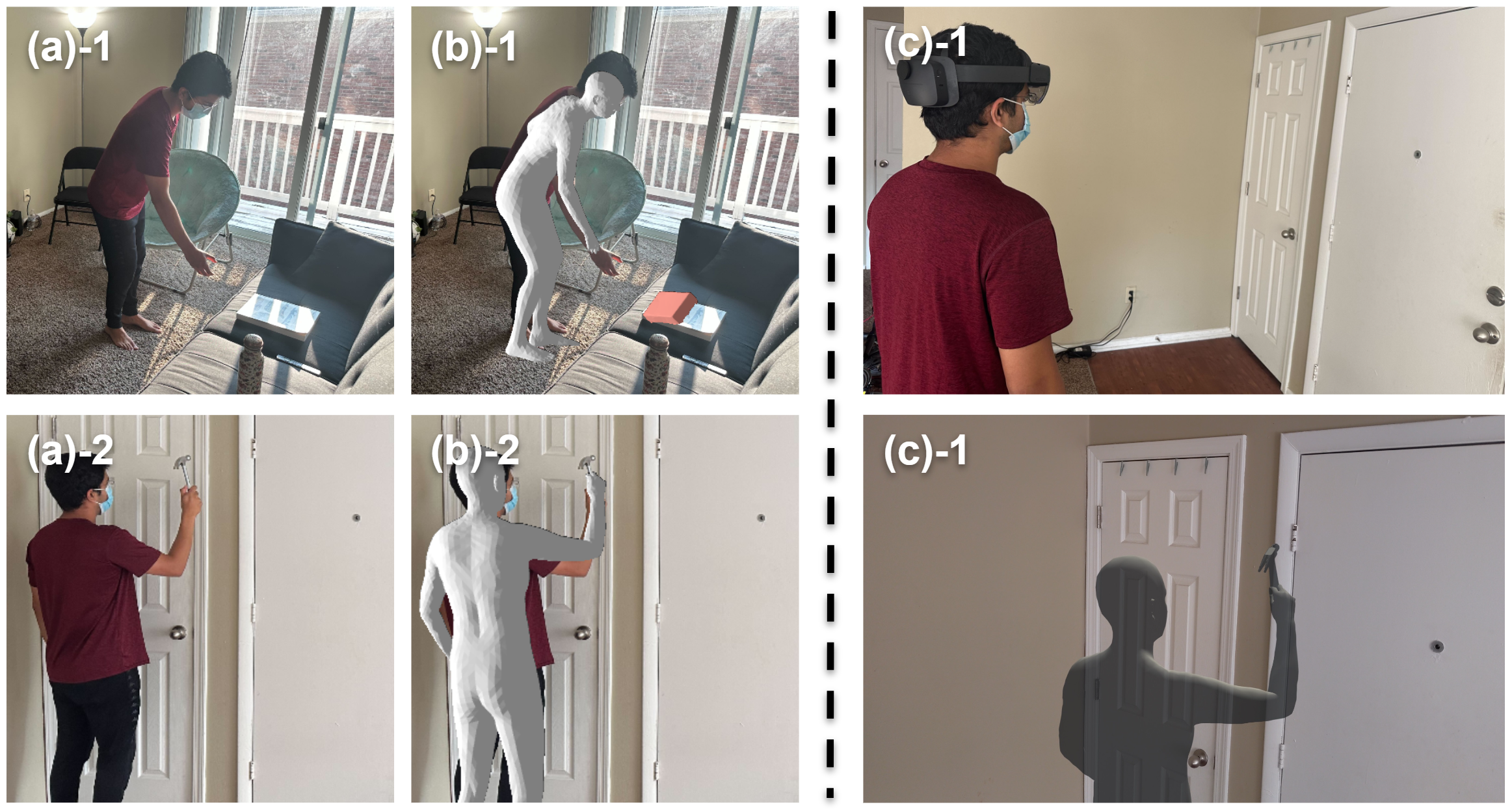}
    \caption{AR animation generated by users using our PbD baselines in User Study 2. (a) users demonstrating the task, (b) the generated 3D animation assets from the camera captures, and (c) the users viewing and adjusting the 3D animation assets with the AR HMD. The differences between the assets in (b) and (c) are due to rendering methods.}
    \label{fig:task_user_study_pbd}
\end{figure*}

With PbD, the users first manually aligned the virtual objects with their real counterparts.
Then, users wrote down the instructions for each task and performed the task in the environment.
During this process, the user's actions were recorded by four camera setups, each capturing one global location in the tasks (sofa, chair, plant, and door).
We followed prior work~\cite{jain2023ubi} to calibrate the camera setups and align them with the AR HMD to obtain accurate camera coordinates.
The recorded videos are then passed into a video-to-3D algorithm~\cite{nam2024contho} to convert the demonstrated motion into presentable 3D humanoid animation assets.
To execute the video-to-3D algorithm, users are first required to segment both the human and the object using the segmentation module from ~\cite{kirillov2023segany}.
The users then situated the animation assets in AR with an HMD, by moving the assets to align with the physical environment.
Until satisfied, the users could redo the tasks and adjust the animation assets.
The examples of generated AR animation with PbD in this study are shown in ~\autoref{fig:task_user_study_pbd}.

For fair comparison of the avatar quality, both PbD and CARING-AI used the full SMPL-X~\cite{pavlakos2019expressive} model as the humanoid avatars as shown in ~\autoref{fig:task_user_study_caring} and ~\autoref{fig:task_user_study_pbd} (c)-1.


\subsection{Results and Analysis} 
We obtained the data from the user study, including (1) the Error Rates (We manually counted the number of times each user modified the instruction, re-performed a task, or re-adjusted the animation assets), (2) the time performance in minutes taken by each user to complete the authoring tasks, and (3) NASA TLX scores.
We then confirmed if the normality assumption is followed in each collected data group with a Shapiro-Wilk test, followed by a paired t-test if normally distributed, or a Wilcoxon Signed-Rank test otherwise.
We then analyzed and discussed the results as follows.

\subsubsection{Task Load: CARING-AI v.s. PbD}
\begin{figure}[ht]
    \centering
    \includegraphics[width=\linewidth]{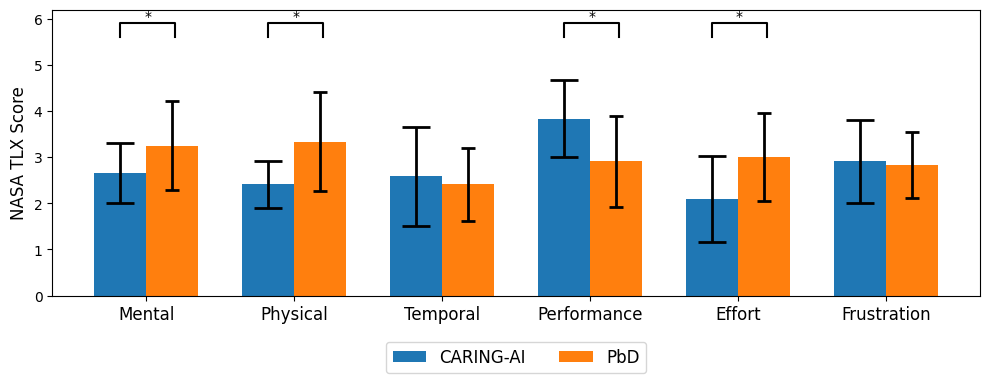}
    \caption{NASA TLX Scores, where * denotes $p < 0.05$ }
    \label{fig:nasatlx}
\end{figure}

Since only data from Effort scores are normally distributed in both PbD and CARING-AI setups ($p_{PbD}=0.051$, $p_{Ours}=0.159$, henceforth, we conducted paired t-tests for Effort and Wilcoxon Signed-Rank tests for the rest, as shown in~\autoref{fig:nasatlx}.
The results showed that users experienced significantly less Mental Demand with CARING-AI ($M_{Ours}=2.666$, $SD_{Ours}=0.651$, $M_{PbD}=3.250$, $SD_{PbD}=0.965$, $p=0.025$,$z=-2.242$) compared to PbD.
Also, users reported significantly higher Physical Demand ($M_{Ours}=2.416$, $SD_{Ours}=0.514$, $M_{PbD}=3.333$, $SD_{PbD}=1.073$, $p=0.046$, $z=-2.001$) in PbD than in CARING-AI.
The less Mental Demand with CARING-AI can be attributed to a shorter workflow with no consideration of the camera position (as we will also discuss in the next subsubsection), while the less Physical Demand with CARING-AI can be attributed to the physical easiness of creating animation with only text instructions compared to that of demonstrating the actions to the cameras.
Additionally, users felt more confident in their performance in completing tasks with CARING-AI ($M_{Ours}=3.833$,$SD_{Ours}=0.834$,$M_{PbD}=2.916$, $SD_{PbD}=0.996$,$p=0.026$,$z=-2.222$).
The better performance scores can relate to the less Error Rates and shorter task time in the next subsubsection.
No significant differences were observed in Temporal Demand ($z=-0.560$,$W=14>W_{critical}=3$) or Frustration ($z=-0.280$,$W=16>W_{critical}=3$) between the two systems.

\subsubsection{Error Rate and Time Performance}
\begin{figure}[ht]
    \centering
    \includegraphics[width=.4\linewidth]{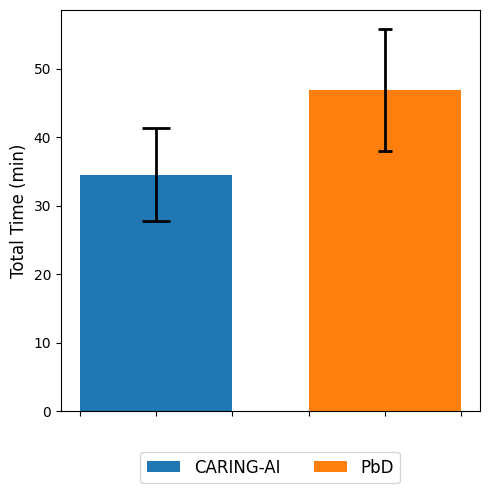}
    \caption{Average Time Spent in Authoring Task, with CARING-AI and PbD}
    \label{fig:error}
\end{figure}

The collected data was normally distributed in both Error Rate ($p_{PbD}=0.515$, $p_{Ours}=0.242$) and Time Performance ($p_{PbD}=0.487$, $p_{Ours}=0.987$).
The reported Error Rates were high in PbD as compared to CARING-AI as shown in~\autoref{fig:error} ($p=0.034$).
During the study, we mainly observed that some participants re-did the tasks with PbD multiple times because the cameras had been occluded from a proper view to generate accurate animation.
In practical scenarios, this problem can worsen since the camera setup has to be relocated and re-calibrated to tackle the occlusion problem of PbD authoring. 
Redoing the demonstration also added much more mental and physical demand as we showed in the NASA TLX results.
For the total time taken, users finished all tasks quicker with CARING-AI compared to PbD  ($p=0.001$).
This was because performing the actions took longer as compared to adjusting the text or the animation itself.
Also, more Error Rates meant more numbers of times re-demonstrating. 


\begin{figure}[ht]
    \centering
    \includegraphics[width=.4\linewidth]{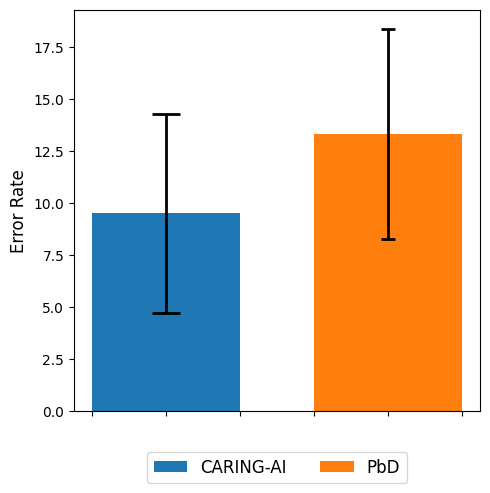}
    \caption{Average Error Rates Calculated in User Study 2, with CARING-AI and PbD}
    \label{fig:slidercount}
\end{figure}

\subsubsection{Subjective Ratings}

\begin{figure*}[ht]
    \centering
    \includegraphics[width=.9\textwidth]{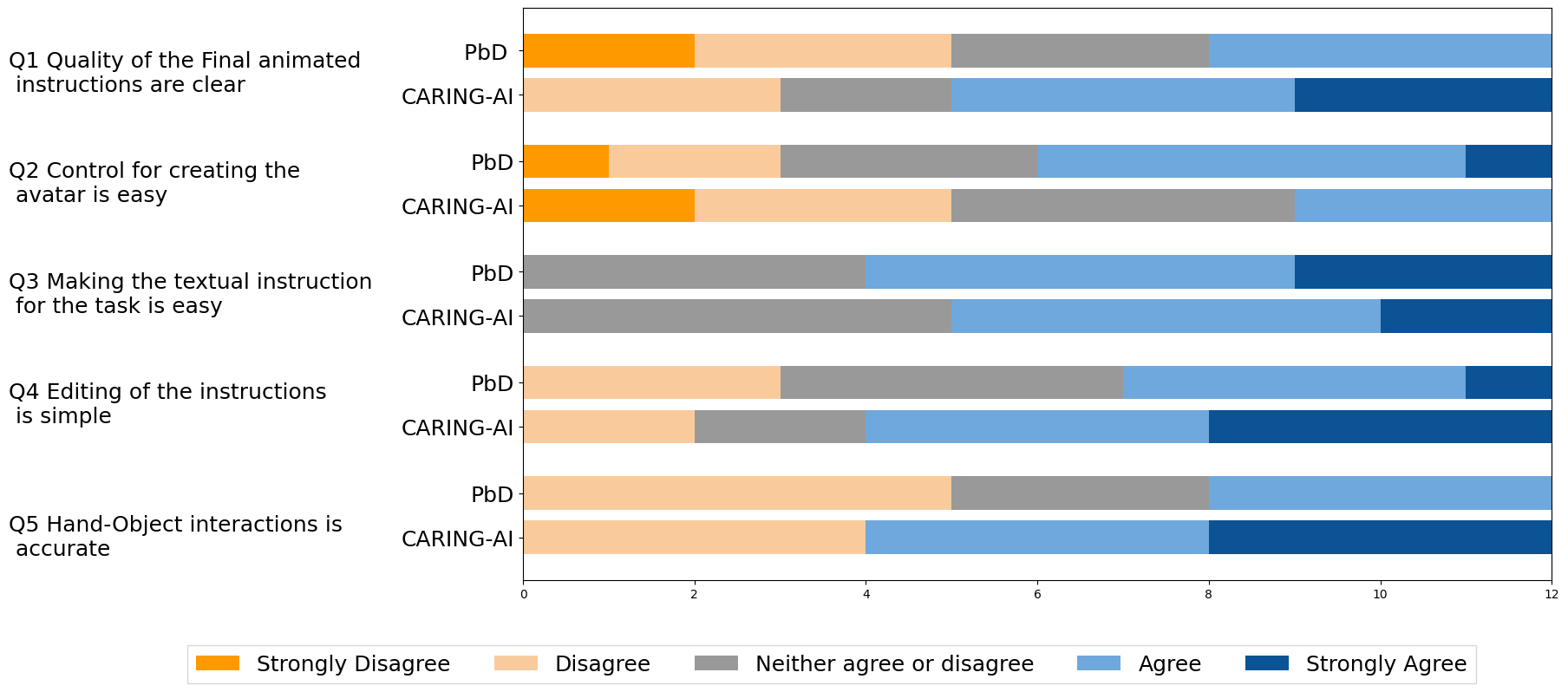}
    \caption{Subjective Likert-scale Ratings of the easiness, animation quality, and interactions of CARING-AI and PbD}
    \label{fig:study2question}
\end{figure*}

We analyzed the questionnaire results from users' feedback and conducted interviews with the participants (\autoref{fig:study2question}).
After confirming the normality of the rating data, we further performed paired t-tests to check the significance of the comparison.
Users preferred the quality of the final animation instruction generated from CARING-AI(Q1: $p < 0.05$), as P7 commented in the interview
\textit{"I like the overall animation quality from the first system (Ours)" (P7).} 
Users found editing animation through text much easier than demonstrating (Q4 $p < 0.05$).
\textit{"When I create the instruction and I don't like it, I would prefer some easier way to edit like the second system (Ours) than performing the task again" (P2).}
This aligns with the results from NASA TLX and quantitative evaluations.
The easiness of authoring with CARING-AI can also be attributed to the smoother learning curve of text-to-animation models as compared to PbD or demonstration-to-animation methods as pointed out in~\cite{shi2023hci}.
Users reported better feedback regarding hand-object rendering in CARING-AI (Q5 $p < 0.05$) as compared to PbD, attributed to the error in detecting hand-object interaction in PbD, which results in the degenerated user experience as described by P3
\textit{"I don't know but the hand was not actually grabbing the object in the first system (PbD) but in the second system It was much better" (P3).}
The quality of hand-object rendering was partially influenced by the occlusion during the study.
Also, the rendered interactions in PbD were not situated in the environment correctly, and users reported that they had adjusted the animations more in the final stage.
We found no significant difference in controlling the avatar (Q2) and editing the textual instruction (Q3). 

\section{Discussion}

\subsection{Contextual Awareness in CARING-AI}
Context plays a vital role in AR applications~\cite{wang2020capturar}. Despite the potential of Gen-AI in creating 3D content~\cite{tevet2022human}, a key limitation lies in its lack of contextual awareness in AR as identified in the preliminary study. One of the core design goals of CARING-AI is to bring context awareness to AI-generated humanoid avatar instructions in AR. 
With the evaluation of the CARING-AI (\autoref{sec:qualeva} and \autoref{sec:quaneva}), we look back at the conclusion drawn in the preliminary study and seek the reason why context-awareness is necessary in AR instruction, i.e., \textit{What does context-awareness bring to AR instructions?}
We highlighted how users in the study emphasized the importance of precise positioning and action performance in humanoid avatar animations (\autoref{sec:caeva}) and demonstrated how CARING-AI effectively addressed the need for accuracy in both positioning and animation.
Our findings indicate that users, as the authors of the instructions, they are aware of the context.
The actions that they intend or anticipate the receivers of the instructions to take are based on the context, i.e., the authors give instructions based on the context.
\textit{"P3: When I wanna instruct somebody to do something, I want them to know exactly the objects and actions. This is very important when in a complex task where students can pick the wrong stuff and act anyway and make mistakes."}
The author's context-awareness is the essence of instructions, as many prior works pointed out as "the prior to function as an instruction" ~\cite{weitzenhoffer1974instruction} or "the flexibility and accommodation to external constraints for designing an instruction"~\cite{morrison2019designing}.
In short, there are specific \textit{"where"} and \textit{"what"} the authors intend to convey in the instructions.
By preserving and representing the author's context awareness, CARING-AI enables the core functions of instructions, which were previously missing in AI-generated humanoid avatar animation.

\subsection{CARING-AI Excels in Instructing: \textit{How}?}
Given the capability of CARING-AI preserving and representing context-awareness in AI-generated humanoid avatar animation. 
We have witnessed the preference and positive ratings of the users for both our full-body avatar and hand-object avatar animation.
Yet, we noticed some comments from the users addressing the necessity of using humanoid avatar instructions in some scenarios.
Some users mentioned inconvenience brought by the use of humanoid avatars.
\textit{"P8: The avatar moving backward was not visible to me. When the movement [of the avatar] is out of my vision, I think there are better ways to tell me to look at the avatar or tell me what to do."} In addition to \textit{"where"} and \textit{"what"}, \textit{"how"} the instructions can be conveyed to the users is also important.
As discovered in prior works ~\cite{cao2020exploratory, huang2021adaptutar}, in AR tutoring, learners prefer half-body avatars for spatial interactions (interactions that require large spatial navigation before proceeding), full-body avatars for body-coordinated interactions (interactions that require coordination among learners' body, hands, and eyes).
We further bring hand-object avatars for local interactions in CARING-AI.
CARING AI changes form of avatar based on scale of the task. However, some of the users mentioned that avatar forms should be based on designation and details of the actions.
Furthermore, as AR instructions are not limited to the form of humanoid avatars, we conclude that non-avatar AR can also be included in the CARING-AI pipeline as a means of visualization.
As P8 commented, non-avatar AR instructions excel avatar instructions in the cases where no particular body gesture is required or the visualization of humanoid avatars is not visible to the users. This conclusion highly aligns with the findings of prior study~\cite{cao2020exploratory}.

\subsection{Other Modalities of AI-generated Instructions}
Humanoid avatar motion along with additional cues helps in learning content~\cite{huang2021adaptutar}. As previously discussed, to further develop CARING-AI into a comprehensive AR instruction system, we envisioned future versions with other AI-generated modalities such as (1) visual cues~\cite{lee2023smartphone, 9874255}, e.g. arrows, bounding boxes, lines, etc., (2) contextualized textual instructions~\cite{chen2023papertoplace}, (3) images~\cite{liu20233dall, radford2021learning}, (4) audio~\cite{louie2020novice}, and (5) videos~\cite{fei2021exposing}.
We argue that our design space of AR instructions and the pipeline of CARING-AI apply to the other modalities of AR cues and instructions as well, since the spatial locations/placements, as well as the temporality of the cues, are key design considerations in the prior works referenced above, and can be situated through CARING-AI's pipeline, where the authors walk through the context and assign the cues by taking contextual snapshots.

\subsection{Authoring with CARING-AI v.s. PbD}
Authoring by real-time demonstration or embodied often requires bulky hardware setup, which limits the mobility of the end-users due to the size of the devices~\cite{wang2020capturar}.
In our User Study 2, a camera setup has to be built for the baseline PbD method, while CARING-AI does not require complex hardware setup and allows users to create instructions without programming knowledge and physical presence.
CARING-AI can help users generate instructions even at remote locations without performing the actions and movements (\autoref{apdx:as3}).
As P7 commented on the efficiency in creating AR instructions \textit{"this system, I think, can become very effective in creating remote instructing like without even physically present."}.
Instructions from PbD are also tied to the environment or context in which users performed demonstrations to create the content. On the other hand, instructions generated from CARING-AI can be effectively adapted to various environments settings because of context-aware modeling, concluded from our observations of the user feedback on the quality of the generated content as shown in~\autoref{fig:study2question}. 
As the results suggest, the CARING-AI pipeline performs better than PbD in generating instructions with fewer mistakes and faster and with less cognitive load.

\section{Limitations, and Future Work}
\label{sec:limitation}
In this section, we discuss the limitations of CARING-AI identified from development, user study, and the analysis of the study process.
Deriving from such, we propose future directions that can contribute to the topic of GenAI in AR instructions.

\subsection{Object Representation and Interactions}

One of the limitations of CARING-AI lies in its ability to handle complex hand-object interactions.
We apply an additional module to render hand-object interactions, which focuses solely on visualizing the hand and object rather than the entire body.
However, this module tackles only rigid objects and does not render high-fidelity hand-object interaction, particularly limiting the use cases requiring complexity and dexterity in manual tasks.
Complex interactions involve hands engaging with objects that are articulated, segmented, foldable, or deformable, whereas the objects in our system are strictly rigid.
Thus, the system cannot represent actions that involve changing the shape or form of an object such as tying a shoelace or folding a cloth.
Nevertheless, such constraints are attributed to the limitation of the Gen-AI algorithms applied in the pipeline, while the overall workflow of CARING-AI remains effective in capturing and presenting the context information to the generated content.
While algorithmic development remains relatively unexplored in the AI field, we foresee this limitation can be addressed in future work by incorporating more generalized state-of-the-art algorithms and datasets, such as those introduced in ~\cite{zheng2023cams, fan2023arctic, lu2023humantomato}, to enable high fidelity rendering of hand-object interactions in AR instruction.

Lastly, CARING-AI currently does not support object-object interactions.
This limitation stems from the aforementioned challenges in hand pose plausibility and rigid object representations.
Without the ability to depict detailed hand-object interactions and object articulations, representing interactions between multiple objects becomes unfeasible.
However, we believe that exploring object-object interactions offers a promising direction for future research, providing a richer and more comprehensive understanding of interactions in virtual environments.

\subsection{Generalizability}
Like all other deep-learning-based methodologies, the performance of our motion generation model is subject to the training process~\cite{justus2018predicting, kalamkar2019study}.
Nevertheless, the model we used has been pre-trained on a large-scale motion dataset~\cite{Guo_2022_CVPR} containing 14,616 motions and 44,970 descriptions composed of 5,371 distinct words, which fulfills the requirement for our use cases and study.
In our study, we emphasize the HCI design and the workflow bridging AR applications and Gen-AI, rather than contributing to the existing algorithms of Gen-AI by trying to outperform them.
To this end, we further argue that the generalizability (more types) and scalability (more detailed motion) of this method are promising.
Firstly, prior works have demonstrated the capabilities of large generative models on large-scale datasets~\cite{rombach2021highresolution}.
We envision the scale of this method will be further improved upon datasets with wider ranges of action labels being fed into the Gen-AI model (e.g. task-specific motion datasets in each domain).
Secondly, the size and complexity of the model in our implementation are constrained by our hardware condition, particularly the GPU sizes.
With a better (empirically more complex) model, we expect the quality and the details of the generated content to improve.

Put simply, our methodology focuses on the HCI design for AIGC in AR, maintains its applicability with the current ideology of Gen-AI, and is generalizable as long as the plugged-in Gen-AI is generalizable.

In addition to the generalizability of the algorithm, we also acknowledge that the findings of the formative studies are derived from academic researchers, which could be further refined and expanded with diverse perspectives from industrial practitioners.
Moreover, the avatars used in our paper are sex-neutral, however, unclothed human avatar representations from~\cite{pavlakos2019expressive}, which can be replaced with inclusive and realistically rendered avatars for more user-friendly and family-friendly use.

\subsection{Software and Hardware Constraints}
One of the major constraints imposed by our hardware condition is the time performance.
It has been reported in ~\autoref{sec:implementation} that generating a batch of motion takes 36 seconds (i.e. anything between 1 and the batch size takes 36 seconds).
Even though 36 seconds of latency seems beyond the cost of real-time performance, batch processing guarantees that users can render their desired avatar instructions once altogether, given a batch size of 128 in our setup, which is, in all cases of our study, more than the users' expectation of the number of interactions in the demonstration.
Moreover, to address this problem of computational cost in the future, we anticipate methodologies such as utilizing cloud services for data transferring and computation, parallel programming for the generation, and usage of better GPUs (high computational).

Another one of our hardware constraints comes from our implementation platform, Hololens 2.
With a field of view (FOV) of (43°×29°), users cannot experience a fully immersive AR environment as content might not be visible outside this boxed area.
For AR authoring and consuming, this poses a challenge.
Users have to be acutely aware of this constraint to ensure that critical interactive elements or information are positioned within this limited space.
Under the circumstances when the humanoid avatar is close to the users, motion outside the FOV is not visible.
This problem may not influence the quality of the generated content itself but induce biases in the evaluation of the user study, such as more negative feedback due to the jeopardized user experience.

\section{Conclusion}
In this work, we present CARING-AI, an AR authoring system that enables users to author AR instructions with contextualized humanoid avatar movement generated by Gen-AI.
We first discussed with experts in AR authoring in a preliminary interview, aiming to identify the gap between current AI-generated humanoid avatars and AR instruction applications.
Based on the insights gained from the discussion, we further characterized the design space for context-aware AR instructions from AI-generated content with two dimensions, namely context (spatial or temporal) and content (local or global).
We then proposed a workflow for contextualizing AI-generated AR instruction with three major steps: (1) generating and modifying textual instructions, (2) contextualizing by traversing and scanning the environment, and (3) generating and smoothing humanoid avatar animation.
We further showcased three application scenarios for authoring AR instructions with CARING-AI: asynchronous, remote, and ad hoc instruction.
We evaluated the performance of CARING-AI with a preliminary quantitative evaluation focusing on the model performance and the quality of the AIGC, followed by a user study evaluating the qualitative performance and overall usability of CARING-AI as an AR authoring system through complimentary qualitative user feedback.
Eventually, we discuss the limitations of the current version of CARING-AI and further envision the opportunities and promising future research directions our work has revealed.
We believe our work is capable of opening up and contributing to the discussion of the broad topic of AIGC in AR applications.

\begin{acks}
We wish to thank all the reviewers for their invaluable feedback.
This work is partially supported by the NSF under the Future of Work at the Human-Technology Frontier (FW-HTF) 1839971.
We also acknowledge the Feddersen Distinguished Professorship Funds and a gift from Thomas J. Malott.
Any opinions, findings, and conclusions expressed in this material are those of the authors and do not necessarily reflect the views of the funding agency.
\end{acks}


\bibliographystyle{ACM-Reference-Format}
\bibliography{reference}

\newpage
\appendix
\section{Application Scenario}

With CARING-AI, users are enabled to author context-aware humanoid avatar animation for AR instructions that can be adaptively deployed into various application scenarios (\textbf{AS}).
Our primary goal of presenting the \textbf{AS} is to demonstrate that CARING-AI can capture and convey context information identified in design space.
Specifically, we showcase three major scenarios where CARING-AI demonstrates its ability to grant code-less and Mocap-free authoring (\textbf{AS-1}, \textbf{AS-2}, and \textbf{AS-3}), create content that is to be deployed in different time primitives or via different platforms (\textbf{AS-1}, \textbf{AS-3}), and adapt to varying contexts (\textbf{AS-2}).

\subsection{AS-1: Asynchronous Instructions}
\begin{figure}[ht]
    \centering
    \includegraphics[width=\linewidth]{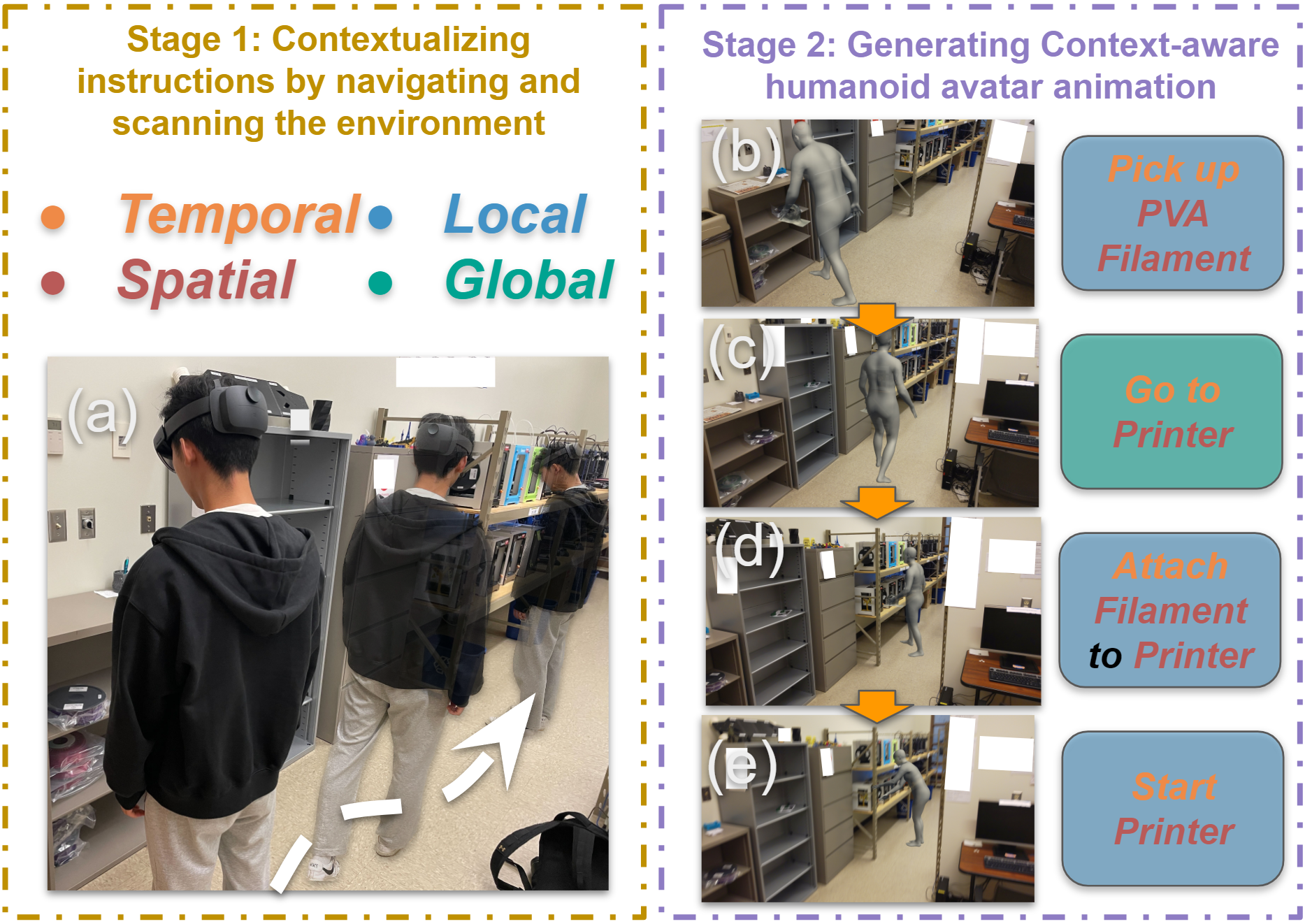}
    \caption{CARING-AI for authoring asynchronous Instructions. A senior lab researcher (a) leaves an AR memo for his colleague on how to use a 3D printer. He simply walks around the printing lab using CARING-AI to contextualize the textual instructions, capturing the location of the PVA filament and the printer. (b) The corresponding humanoid animation is generated according to the step-by-step instructions. CARING-AI is capable of handling AR instructions of diverse content and context, namely spatial or temporal context, and local or global content.}
    \label{fig:usecase1}
\end{figure}
\vspace{-0.5cm}
Asynchronous instructions are the most common case in the applications of AR instructions, where the authors create the content prior to the consumption of the AR experiences~\cite{chen2023papertoplace, chidambaram2021processar}.
CARING-AI naturally supports asynchronous instructions and situates AI-generated humanoid avatar animation into the physical world contextually.
Here, we showcase a scenario in a research lab, where a senior researcher (Tom, the author) would like to leave an AR memo for his junior colleague (Jerry, the consumer) to instruct him on how to operate a 3D printer.
Tom creates and modifies the text instructions with the help of CARING-AI, then provides context to the system by walking up to the locations and taking snapshots of the environment as shown in ~\autoref{fig:usecase1}.
He informs Jerry to get the printing materials and then go to a specific 3D printer to print a product.
Later, when Jerry arrives in the laboratory, he follows the step-by-step AR memo from Tom to start working on the product.

\begin{figure}[htp]
    \centering
    \includegraphics[width=\linewidth]{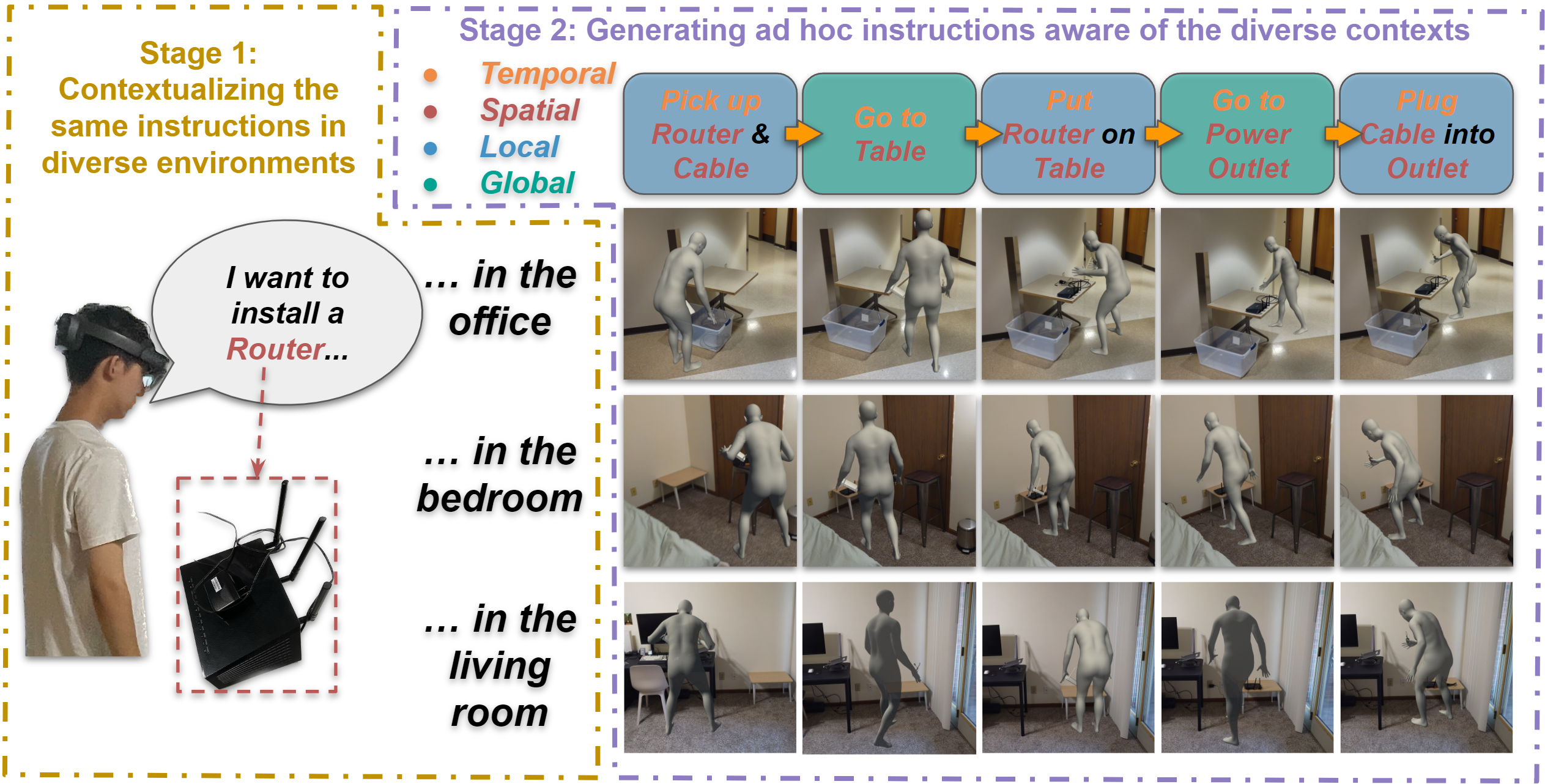}
    \caption{CARING-AI for Ad Hoc AR Instructions. For the same task (e.g. installing a router), the instructions vary across diverse contexts. By using CARING-AI, users simply need to scan the environment to provide contextual information to the system. CARING-AI will generate humanoid avatar animation that blends into different physical realities.}
    \label{fig:usecase2}
\end{figure}

\subsection{AS-2: Ad Hoc Instruction Creation}

CARING-AI enables authoring AR instructions through Gen-AI by contextualizing the instructions.
In this scenario, we showcase how CARING-AI enables authoring ad hoc instructions in changing contexts with simplified user interactions.
As a technician from the lab, Tom would like to teach his colleague how to install a router as shown in ~\autoref{fig:usecase2}.
The instructions are fairly simple and easy to understand.
However, the detail of the steps varies across environments, e.g. in the office, the bedroom, or the living room, because the locations of the router and the outlet vary.
With the same protocol to be visualized, Tom simply has to contextualize the protocol in different places, assigning the locations of the objects by traversing the rooms.
As a result, Tom authors different avatar animations for diverse contexts with the same instruction protocol.

\begin{figure}[htp]
    \centering
    \includegraphics[width=\linewidth]{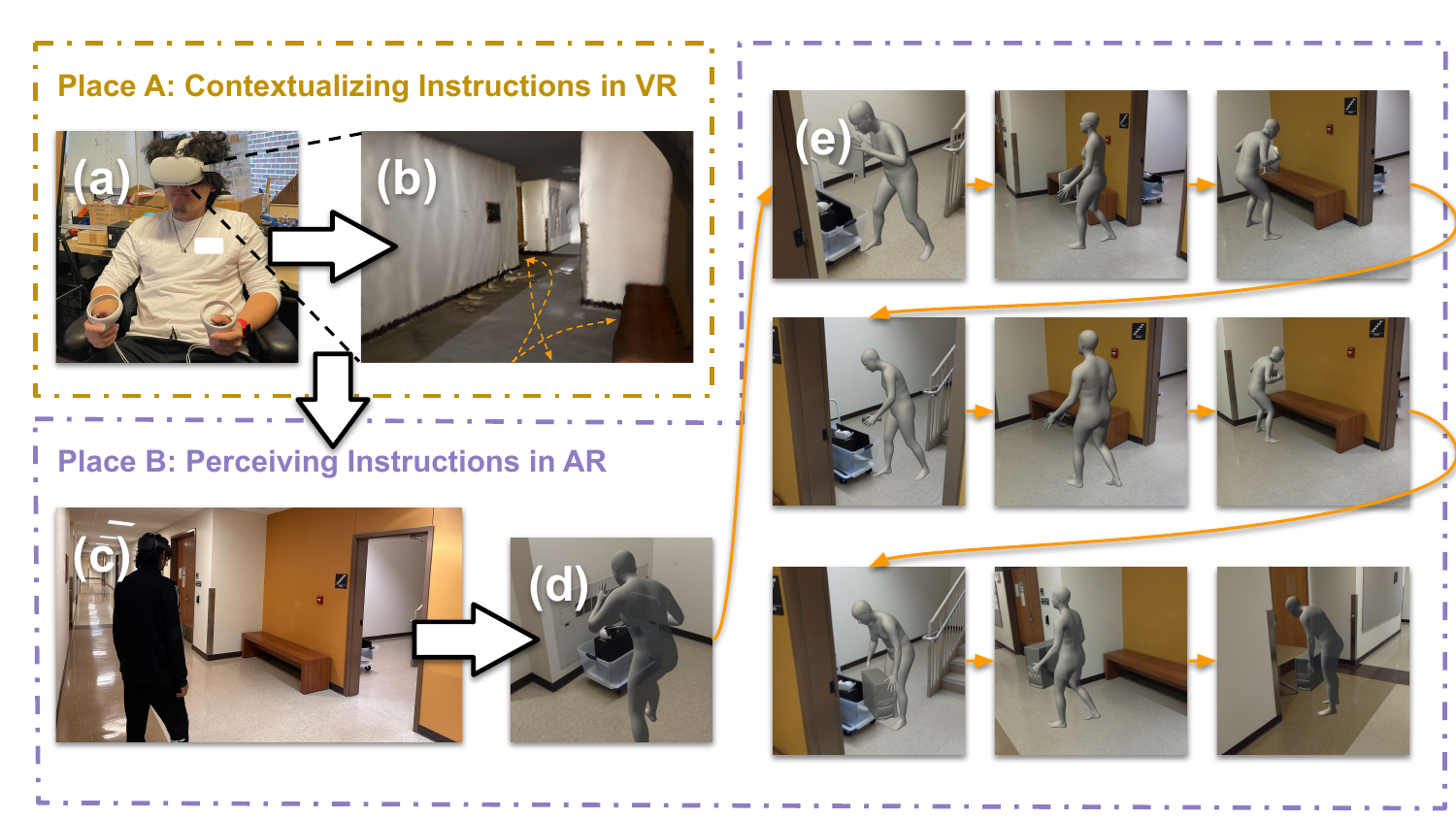}
    \caption{CARING-AI's capability of authoring Remote Instructions. In this scenario, a delivery man is asking for the destinations of the packages. (a) A lab member is giving contextual information through a pre-scanned scenario in VR. (b) We built a mock-up VR scenario to record the locations and correspond them back into the physical reality. Once the instructions are contextualized, the delivery man can view the humanoid instructions on delivering the packages (c, d).}
    \label{fig:usecase3}
\end{figure}

\begin{figure*}[htp]
    \centering
    \includegraphics[width=\linewidth]{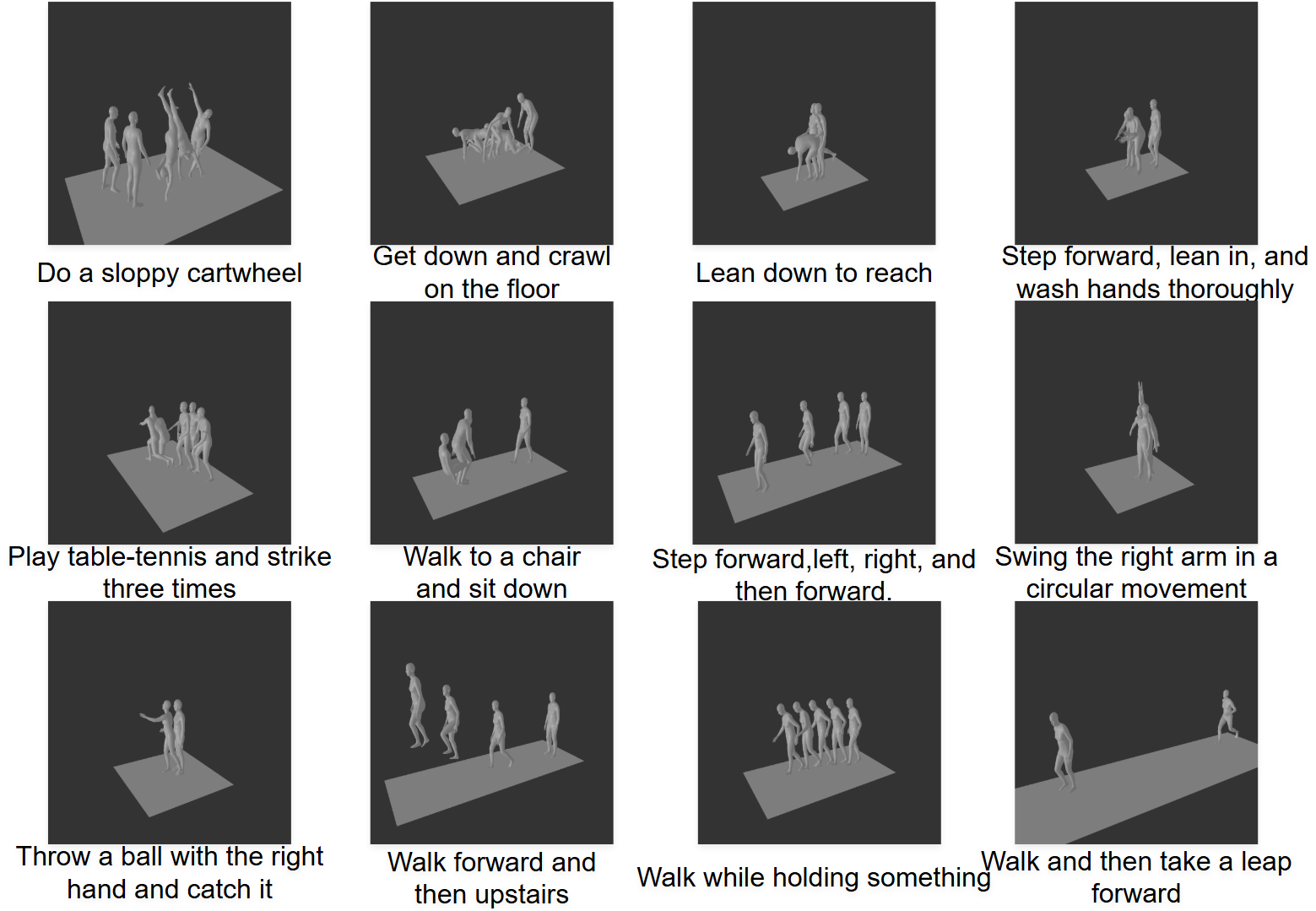}
    \caption{We showcase more examples of humanoid animation generated from our backend algorithm.
    Specifically, the animations generated are guided by a textual description of the motion, rendered with humanoid avatars (in our implementation, SMPL~\cite{pavlakos2019expressive}).
    Each animation clip presents a short sequence of human motion and represents a step in a given AR instruction.}
    \label{fig:apdx_animation}
\end{figure*}

\subsection{AS-3: Remote Instructions}
\label{apdx:as3}
In this scenario, CARING-AI is deployed in a remote instruction task.
We showcased how CARING-AI can adapt to context information of diverse modalities and liberate the authors from demonstrating in the actual physical environment.
Toodles, the deliverywoman of the building, arrives in the lab with new devices to be allocated ~\autoref{fig:usecase3} (c).
Noticing no one is in the lab, Toodles contacts Jerry, asking about the allocation of the devices.
Jerry, who is not present at the lab, confirms the devices and their checkout points (i.e. where they are to be placed).
Jerry then enters a pre-scanned point-cloud map of the lab in Virtual Reality (VR), where he authors the instructions in VR using CARING-AI by navigating the map and taking screenshots ~\autoref{fig:usecase3} (a, b) (We built a mock-up VR program to record Jerry's locations in VR and correspond them to the physical reality).
CARING-AI generates humanoid avatar instructions according to the contextual information provided.
The authored AR instructions are then sent to Toodles, who follows the avatar demonstrations to allocate the devices to different locations~\autoref{fig:usecase3} (d, e).
In this case, we see that CARING-AI is capable of authoring synchronous remote instructions.
It also showcases the possibilities of authoring AR experiences in VR with CARING-AI with aligned contextual information between physical reality and VR. The alignment of context is subsumed here as described and inspired by many prior works~\cite{qian2022scalar, Nebeling2020XRDirector, wang2021distanciar}

\section{More Generated Examples}
\begin{figure*}[htp]
    \centering
    \includegraphics[width=\linewidth]{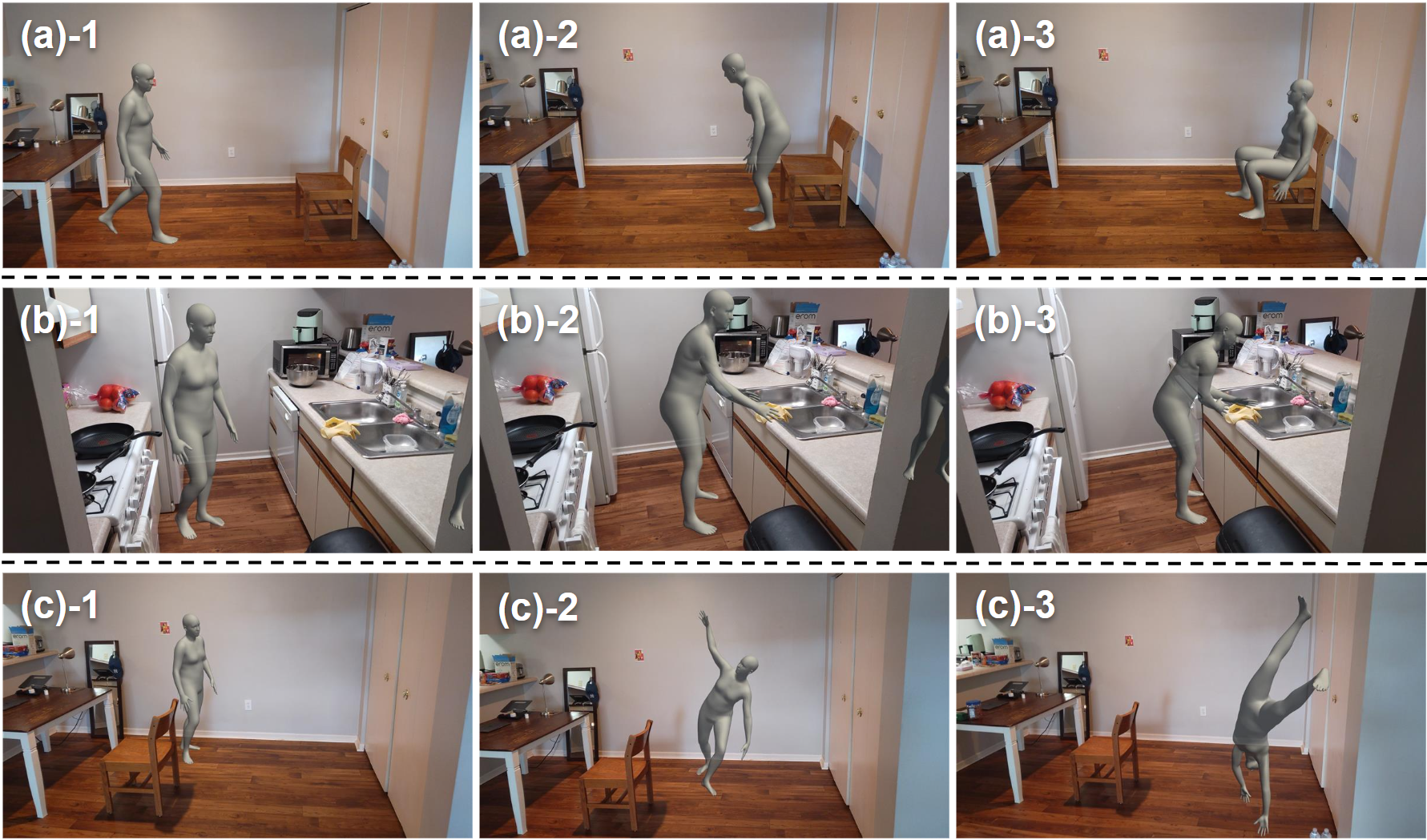}
    \caption{We showcase more examples of humanoid animation contextualized by CARING-AI, (a) Walk to the chair and sit down, (b) Walk to the sink, lean in, and wash hands thoroughly, and (c) do a a sloppy cartwheel around the chair. All animations are generated by prompting the CARING-AI with text, scanning the environment to mark the object, and passing user trajectories to the generative model.} 
    \label{fig:apdx_ar}
\end{figure*}

In this section, we showcase more examples generated from our backend diffusion model as shown in ~\autoref{fig:apdx_animation} and situated humanoid animation by CARING-AI through our pipeline as shown in ~\autoref{fig:apdx_ar}.
Given a textual prompt, our motion diffusion model can generate high-fidelity humanoid avatar motion.
With the user-provided context, specifically object location and motion trajectory, the CARING-AI system can situate the generated animations in the space and temporally smooth them for a seamless user experience.

\appendix

\end{document}